\documentclass{jpp}
\usepackage{graphicx}
\usepackage{url}
\usepackage[utf8]{inputenc}
\usepackage[T1]{fontenc}
\usepackage{amsmath}
\usepackage{graphicx}
\usepackage{babel}
\usepackage{txfonts}
\usepackage[normalem]{ulem}

\usepackage{soul}
\usepackage{xcolor}
\usepackage{amsmath}
\DeclareMathOperator{\Tr}{Tr}
\usepackage[mathscr]{euscript}

\usepackage{lineno}

   \title{Role of FLR effects in magnetopause equilibrium}
   \author{G. Ballerini
            \corresp{\email{giulio.ballerini@lpp.polytechnique.fr}}
          \aff{1,2}
          \and
          L. Rezeau\aff{1}
          \and
          G. Belmont\aff{1}  
          \and
          F. Califano\aff{2}
          }

   \affiliation{\aff{1}LPP, CNRS/Sorbonne Université/Université Paris-Saclay/Observatoire de Paris/Ecole Polytechnique, Institut Polytechnique de Paris, Paris, France
         \aff{2}Dipartimento di Fisica E. Fermi, University of Pisa, Italy
             }

\begin{document} 
\maketitle
 
  \abstract{
   
    The Earth magnetopause, when sufficiently plane and stationary at a local scale, can be considered as a "quasi-tangential" discontinuity, since the normal component of the magnetic field $B_n$ is typically very small but not zero. Contrary to observations, the "Classic Theory of Discontinuities" (CTD) predicts that rotational and compressional jumps should be mutually exclusive in the general case $B_n \ne 0$, but allows only one exception: the tangential discontinuity provided that $B_n$ is strictly zero. Here we show that Finite Larmor Radius (FLR) effects play an important role in the quasi-tangential case, whenever the ion Larmor radius is not fully negligible with respect to the magnetopause thickness. By including FLR effects, the results suggest that a rotational discontinuity undergoes a change comparable to the change of a Shear Alfvén into a Kinetic Alfv\'en wave when considering linear modes. For this new kind of discontinuity,  the co-existence of rotational and compressional variations at the magnetopause does no more imply that this boundary is a strict tangential discontinuity, even in 1D-like regions far from X-lines if any. This result may lead to important consequences concerning the oldest and most basic questions of magnetospheric physics: how can the magnetopause be open, where and when?
    The role of FLR being established theoretically, the paper then shows that it can be proved experimentally. For that, we make use of MMS data and process them with the most recent available 4 spacecraft tools. First, we present the different processing techniques that we use to estimate spatial derivatives, such as $grad(B)$ and $div(P)$, and the magnetopause normal direction. We point out why this normal direction must be determined with extremely high accuracy to make the conclusions unambiguous. Then, the results obtained by these techniques are presented in a detailed case study and on a statistical basis.
    

   \maketitle
%

\section*{Introduction}
\noindent In space physics, there is a natural tendency of the medium to self-organize into distinct cells, separated by thin layers. This behavior can be observed at very different scales. Notable examples are planetary magnetospheres, which are bubbles in the solar wind stream and which are separated from it by bow shocks and magnetopauses ~\citep{parks_physics_2019, kivelson_introduction_1995, belmont_collisionless_2014}. The interaction of the solar wind with unmagnetized bodies such as comets also produces similar bubbles ~\citep{coates_ionospheres_1997, bertucci_structure_2005}. The Solar System itself is a bubble in the flow of the local interstellar cloud, and it is separated from it by the heliopause and at least one shock ("termination shock") ~\citep{lallement_2001, richardson_observations_2022}. Similar cells and thin layers can also form spontaneously, far from any boundary condition as in the context of a turbulent medium~\citep{frisch_turbulence_1995,chasapis_thin_2015}.

Among all these thin layers, the terrestrial magnetopause plays a particular role. This region has been explored by a large number of spacecraft since the beginning of the space era, up to the most recent multi-spacecraft missions as Cluster \citep{Escoubet1997, Escoubet2001} and MMS \citep{burch_phan_2016}, allowing for a detailed description of its properties. In addition, due to a very small normal component of the magnetic field with respect to the magnetopause (defined $B_n=\mathbf{B}\cdot\mathbf{n}$ where $\mathbf{B}$ is the magnetic field and $\mathbf{n}$ the magnetopause's normal) it can be identified as a "quasi-tangential" layer. This feature is a direct consequence of the frozen-in property that prevails at large scales, on both sides of the boundary, almost preventing any penetration of magnetic flux and matter between the solar wind and the magnetospheric media (both of them being magnetised plasmas). By large scales here we refer to the fluid scales where an ideal Ohm's law holds, as in the ideal magnetohydrodynamic (MHD) regime. However, small departures from a strict separation between the two plasmas do exist, at least locally and for a given time interval, and they are known to have important consequences for all the magnetospheric dynamics: substorms, auroras, etc~\citep{mcpherron_magnetospheric_1979, tsurutani_interplanetary_2001}.

Knowing when and where plasma injection occurs through the magnetopause has been one of the hottest subjects of research since decades (\citet{haaland_20_2021} and references therein, \citet{Lundin1984,Gunell2012,Paschmann2018}). The largest consensus presently considers the equilibrium state of the boundary, valid on the major part of its surface, as a tangential discontinuity, with a strictly null $B_n$, while plasma injection is allowed only around a few reconnection regions, where the gradients characterizing the layer present 2D features. For that purpose, many studies have been carried out to understand where magnetic reconnection occurs the most ~\citep{fuselier_antiparallel_2011, trattner_location_2021}. Moreover, the conditions under which the magnetopause opens due to magnetic reconnection has been studied theoretically ~\citep{swisdak_diamagnetic_2003} and experimentally ~\citep{gosling1982, Paschmann1984, phan_extended_2000, fuselier_antiparallel_2011, vines2015}.  The results of the present study may allow reconsidering this paradigm by questioning the necessity of a strictly tangential discontinuity for the basic equilibrium state.\\

\noindent In the whole paper hereafter, we will call one-dimensional all geometries in which the gradients of all parameters are in the same direction $\mathbf N$. In this sense, a plane magnetopause with not tangential gradient is said here to be 1D, while it would be considered 2D if considering real space instead of $k$ space. 

\section{Classic Theory of Discontinuities}

At every layer, the downstream and upstream physical quantities are linked by the fundamental conservation laws: mass, momentum, energy and magnetic flux \citep{landau_fluid}. The simplest case occurs whenever the number of conservation laws is equal to the number of parameters characterizing the plasma state. When this condition is met, the possible downstream states are uniquely determined as a function of the upstream state, regardless of the (non-ideal) physics at play within the layer. In particular, it is possible to describe pressure variations without any closure equation. In this case, the jumps of all quantities are determined by a single scalar parameter (namely the "shock parameter" in neutral gas).

We refer hereafter to the "Classic Theory of Discontinuities" (CTD) as for the theory corresponding to this condition, which is used both for neutral media and (magnetized) plasmas. CTD is characterized by the following simplifying assumptions: a stationary layer, 1D variations, and isotropic pressure on both sides. For plasmas, the additional assumption of an ideal Ohm's law on both sides is considered ~\citep{belmont_introduction_2019}.

In CTD the conservation laws provide a system of jump equations between the upstream and downstream physical quantities, namely the Rankine-Hugoniot conditions in neutral media and generalized Rankine-Hugoniot conditions in plasmas.  

The sets of equations used to compute the linear modes in hydrodynamics (HD) and MHD are similar to these jump equations system. simply because the HD and MHD models rely on the same conservation laws as Rankine-Hugoniot and generalized Rankine-Hugoniot respectively A direct consequence is that many properties are shared by the solutions of the two types of systems: linear modes and discontinuities. For a neutral medium, the linear sound wave solution corresponds to the well-known sonic shock solution, while for a magnetized plasma, the two magnetosonic waves correspond to the two main types of MHD shocks: fast and slow. However, an additional discontinuity solution, the intermediate shock, has no linear counterpart. The intermediate shock presents a reversal of the tangential magnetic field through the discontinuity, which is not observed neither in the fast nor in the slow mode. Furthermore, a non-compressional solution exists in both types of systems, represented by the shear Alfvén mode for linear MHD, and by the "rotational discontinuity" solution for the generalized Rankine-Hugoniot system.

Focusing on magnetized plasma physics, CTD leads to distinguish compressive and rotational discontinuities. An important feature of these solutions is that the compressional and rotational solutions are mutually exclusive: the shock solutions are purely compressional, without any rotation of the tangential magnetic field (this is called the "coplanarity property"), while the rotational discontinuity does imply such a rotation but without any variation of the magnetic field amplitude and without any compression of the particle density (Fig.{\ref{fig:cartoon})}. This distinction persists whatever the fluxes along the discontinuity normal, even when the normal components $u_n$ and $B_n$ of the velocity and the magnetic field are arbitrarily small. The only exception is the "tangential discontinuity" when both normal fluxes are strictly zero. This solution would correspond, for the magnetopause, to the case without any connection between solar wind and magnetosphere. It appears as a singular case since the tangential discontinuity, with $B_n=0$, is not the limit of any of the general solutions with $B_n \ne 0$. While the limit always implies two solutions, one purely rotational and the other purely compressional, the singular solution $B_n=0$ only provides one solution where the two characters can coexist. 


In the solar wind, discontinuities are routinely observed and several authors have performed statistics for a long time to determine the proportion of the different kinds of discontinuities, mainly focusing on the tangential and rotational ones. They conclude that in most cases tangential discontinuities ($i.e.$ with $B_n$ small enough to be barely measurable) are the most ubiquitous (see ~\citet{Colburn1966}, for a pioneering work in this domain and ~\citet{neugebauer_comment_2006, paschmann_discontinuities_2013, Liu2022}, and references therein, for more recent contributions). In these studies, rotational discontinuities are identified only when $B_n$ is large enough. However, many discontinuities present features that are typical of both rotational and tangential discontinuities and are classified as "either" of the two. Extending these studies in the range of small $B_n$, where all discontinuities are not necessarily "tangential discontinuities" in the CTD sense, requires the study of the quasi-tangential case. 

\section{The Earth's magnetopause}
Thanks to \textit{in-situ} observations, the Earth's magnetopause has a pivotal role in testing the discontinuity theories. Indeed, the Earth's magnetopause boundary exhibits, over its entire surface,  both a rotation of the magnetic field ~\citep{sonnerup_magnetopause_1974} and a density variation~\citep{Otto2005} since it is the junction of two media, the magnetosheath and the magnetosphere where the magnetic field and the density are different ~\citep{dorville_rotationalcompressional_2014}.

As stated above, the usual paradigm is that the magnetopause is always a tangential discontinuity and that it becomes "opens" only exceptionally at a few points where the boundary departs from one-dimensionality due to magnetic reconnection. Does it mean that it justifies the very radical hypothesis of a magnetopause nearly completely impermeable to mass and magnetic flux, with strictly null $B_n$ and $u_n$ and quasi-independent plasmas on both sides (apart from the normal pressure equilibrium)? From a theoretical point of view, it is clear that the singular limit from $B_n \simeq 0$ to  $B_n=0$ remains to be solved. From an experimental point of view, if the components $B_n$ and $u_n$ are known to be always very small, the observations can hardly distinguish between $B_n \simeq 0$ and $B_n=0$ because of the uncertainties, due to the fluctuations and the limited accuracy in determining the normal direction ~\citep{Rezeau2017,haaland2004, dorville_magnetopause_2015}.

The results of the present paper will question the above paradigm. We will show theoretically and experimentally that CTD fails at the magnetopause and that rotation and compression can actually coexist with finite $B_n$ and $u_n$, even in the 1D case. Such a paradigm change may be reminiscent of a similar improvement in the theoretical modeling of the magnetotail in the 70's studies (\citet{coppi_dynamics_1966, galeev_reconnection_1979, coroniti_tearing_1980} and references therein). In that case the authors demonstrated that even a very weak component of the magnetic field across the current layer was sufficient to completely modify the stability properties of the plasmasheet, so that the finite value of $B_n$ had to be taken into account, contrary to the pioneer versions of the tearing instability theories.   

\section{The role of pressure}
\begin{figure}
    \centering
    \includegraphics[width=\columnwidth]{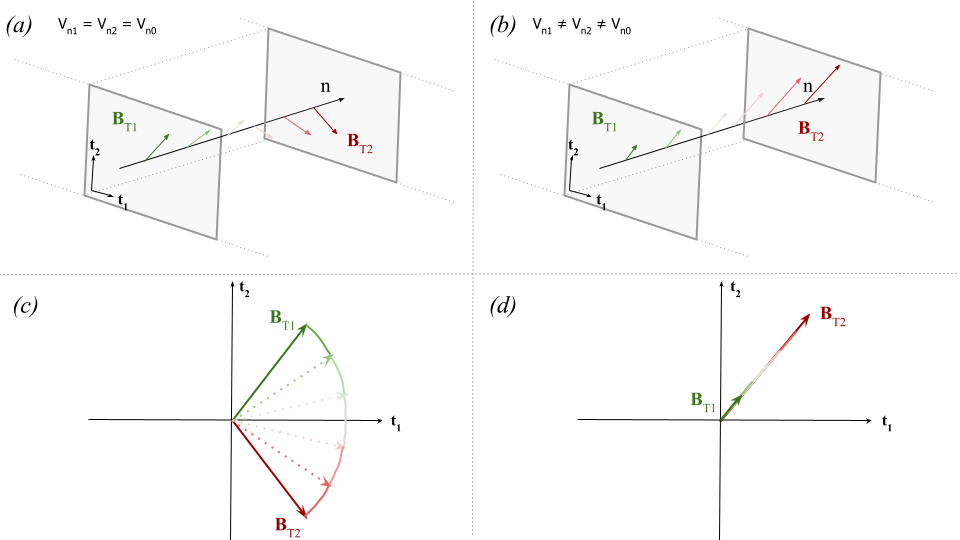}
    \caption{Cartoon showing the different variations of $B$ between a rotational discontinuity (left) and a compressive one (right). The top panel shows in 3D the variation of $B$  inside the magnetopause plane; the bottom panel shows the hodogram in this tangential plane: a circular arc for the rotational discontinuity and a radial line for shocks.}
    \label{fig:cartoon}
\end{figure}
In CTD the separation between the compressional and rotational properties of the discontinuities comes from only two equations projected on the tangential plane. These equations are the momentum equation and the Faraday - Ohm's law, that read:
\begin{equation}
\rho\frac{d \mathbf{u}}{d t}+\nabla\cdot\mathbf{P}_i+\nabla\cdot\mathbf{P}_e=\mathbf{J}\times \mathbf{B}
\label{eq:momentum}
\end{equation}
\begin{equation}
    \mathbf{\nabla}\times \mathbf{E}=-\frac{\partial \mathbf{B}}{\partial t} \nonumber
\end{equation}
\begin{equation}
    \text{where \;\;\;    } \mathbf{E}=-\mathbf{u}\times\mathbf{B}+\frac{1}{ne}\mathbf{J}\times\mathbf{B}-\frac{1}{ne}\nabla\cdot\mathbf{P}_e
\end{equation}
where $\mathbf B$ is the magnetic field and $\mathbf u$ is the flow velocity in a reference frame where the layer is steady.

Considering one-dimensional gradients along the normal direction $\mathbf n$, neglecting the non-ideal terms in Ohm's law and integrating across the layer, these two equations, projected on the tangential plane, give:
\begin{equation}
\rho_{2} u_{n2} \, \mathbf u_{t2}-B_{n2} \, \mathbf B_{t2}/\mu_0 = \rho_{1} u_{n1} \, \mathbf u_{t1}-B_{n1} \, \mathbf B_{t1}/\mu_0
\label{eq:momentum2}
\end{equation}
\begin{equation}
B_{n2} \, \mathbf u_{t2}-u_{n2} \, \mathbf B_{t2}=B_{n1} \, \mathbf u_{t1}-u_{n1} \, \mathbf B_{t1} 
\label{eq:3.4}
\end{equation}

Due to the divergence free equation, the values $B_{n1}$ and $B_{n2}$ are equal and will be written as $B_{n}$ without index in the following. Similarly, $\rho_1 u_{n1}$ and $\rho_2 u_{n2}$ are equal because of the continuity equation and will be simply noted $\rho u_n$ in the following. Here, the indices $n$ and $t$ indicate the projection along the normal and in the tangential plane, respectively, while indices 1 and 2 indicate the two sides of the discontinuity. It is important to note that, in CTD, the pressure divergence terms do not appear in Eq.(\ref{eq:momentum2}) because of the assumption done in this theory that the pressure is isotropic on both sides so that their integration gives terms of the form $(p_2-p_1)\mathbf n$, with no component in the tangential plane.

We see that all terms in these two equations are proportional to $B_n$ or $u_n$, so that any non-ideal term, even small, can become dominant 
when these two quantities tend to zero (if these non-ideal terms do not tend to zero at the same time). As the distinction between compressional and rotational character fully relies on this system of equations, this evidences the necessity of investigating the quasi-tangential case for resolving the usual singularity of the tangential discontinuity. 
We note that the LHS and RHS of equation \ref{eq:3.4} can be put equal to zero by choosing the "De Hoffmann-Teller" tangential reference frame where the electric field is zero \citep{belmont_introduction_2019}. However, this choice, even if it can simplify some calculations, is not necessary here. Finally, the variables $\mathbf u_{t}$ can be eliminated from the system by a simple linear combination of the two equations, leading to:
\begin{equation}
(u_{n2} -u_{n0})\mathbf{B}_{t2} = (u_{n1} -u_{n0})\mathbf{B}_{t1} 
    \label{eq:isotropic}
\end{equation}
where
\begin{equation}
   u_{n0}=\frac{B_{n}^2} {\mu_0 \rho u_n} = \mathrm{cst}  
\end{equation}

Equation (\ref{eq:isotropic}) leads to the distinction between shocks, where the tangential magnetic fields on both sides are collinear (but with different modules), and rotational discontinuities, where the terms inside the brackets must be equal to zero. Rotational discontinuities correspond to a propagation velocity equal to the normal Alfv\'en velocity, and imply $u_{n1}=u_{n2}=u_{n0}$, and therefore an absence of compression of the plasma.

As previously stated, the separation between the compressional and rotational characters mainly derives, in CTD, from the assumption of isotropic pressures on both sides, which prevents the pressure divergences to have tangential components. 
When the isotropic hypothesis is relaxed ~\citep{hudson_rotational_1971}, the set of conservation equations is no longer sufficient to determine a unique downstream state for a given upstream one. As a consequence, the global result depends on the non-ideal processes occurring within the layer. In addition to anisotropy effects, Finite Larmor Radius (FLR) effects can be expected to break the gyrotropy of the pressure tensor around $\mathbf B$ in the case of thin boundaries between different plasmas. This means that the main effect that explains departures from CTD comes from the tangential component of the divergence of the pressure tensor, which must be taken into account in the momentum equation. On the other hand, the non-ideal effects related to the generalized Ohm's law are negligible, at least in the examples shown in this paper.
The possible types of discontinuities in an anisotropic plasma have been discussed in several papers long time ago~\citep{lynn_discontinuities_1967,abraham-shrauner_propagation_1967, chao_interplanetary_1970, neubauer_jump_1970}, and the present paper improves the analysis in the light of the new experimental possibilities given by the MMS measurements. 

When the dynamics drives the conditions for the pressure tensor to become anisotropic (and \textit{a fortiori} in the non-gyrotropic case) the $\nabla\cdot\mathbf{P}$ term comes into play linking upstream and downstream quantities. Considering the "simple" anisotropic case, $i.e.$ keeping the gyrotropy around $\mathbf B$, it has been shown ~\citep{hudson_rotational_1971} that the $\nabla \cdot \mathbf{P}$ term then just introduces a new coefficient:
\begin{equation}
    \alpha = 1-\frac{p_{\parallel}-p_{\perp}}{ B^2 / \mu_0}
    \label{alpha}
\end{equation}

This coefficient has been interpreted as a change in the Alfv\'en velocity $V'^2_{An}=\alpha V_{An}^2$, but it appears more basically as a change in Eq.(\ref{eq:isotropic}):
\begin{equation}
    (u_{n2} -\alpha_2 u_{n0})\mathbf{B}_{t2} = (u_{n1} -\alpha_1 u_{n0})\mathbf{B}_{t1}
    \label{eq:anisotropic}
\end{equation}

This equation shows that, in this simple anisotropic case, coplanar solutions still exist ($\mathbf{B}_{t2}$ and $\mathbf{B}_{t1}$ are collinear), but that whenever $\alpha_2$ is not equal to $\alpha_1$, the equivalent of the rotational discontinuity now implies compression:
\begin{equation}
    u_{n2}  \ne u_{n1}  \quad \quad \text{if} \quad \quad \alpha_2 \ne \alpha_1
    \label{eq:rotational_anisotropic}
\end{equation}
Since $u_{n2}=\alpha_2 u_{n0}$ and $u_{n1}=\alpha_1 u_{n0}$. The variation of $u_n$ explains why the modified rotational discontinuity can be "evolutionary" \citep{jeffrey_non-linear_1964}, the non linear steepening being counter-balanced at equilibrium by non-ideal effects for a thickness comparable with the characteristic scale of these effects.

There is actually no additional conservation equation available that would allow the jump of the anisotropy coefficient $\alpha$ to be determined. Consequently, there is no universal result that gives the downstream state as a function of the upstream one, regardless of the microscopic processes going on within the layer. This remains valid for the full anisotropic case, with non-gyrotropy. As soon as the ion Larmor radius $\rho_i$ and the ion inertial length $d_i$ are not fully negligible with respect to the characteristic scale $L$ of the layer, kinetic effects, and in particular FLR effects, which make the pressure tensor non-gyrotropic, must be taken into account to describe self-consistently the internal processes. Then, the effect of the divergence of the pressure tensor is no longer reduced to adding a coefficient $\alpha$ since its tangential component is no longer collinear with $\textbf B_t$. Such effect has been already reported and analyzed in the context of magnetic reconnection ~\citep{aunai2013, aunai_proton_2011} and in kinetic modeling of purely tangential layers~\citep{belmont_kinetic_2012, dorville_asymmetric_2015}. It has been also investigated in the case of linear modes where they are responsible for the transition from shear Alfvén into Kinetic Alfv\'en Wave ~\citep{Hasegawa,belmont_rezeau1987,cramer}. On the other hand, it has never been introduced in the context of quasi-tangential discontinuities.

If a simple anisotropy preserving gyrotropy around $\mathbf{B}$ can be straight fully taken into account for modeling the pressure tensor and using it in fluid equations, introducing non-gyrotropy does not lead to a general and simple modeling for the pressure tensor. It would demand \textit{a priori} a full kinetic description or, at least, some expansions assuming that these effects are small enough (see \citet{braginskii_transport_1965} for the pioneer work in this field and \citet{passot_fluid_2006} and references therein). Several papers have investigated the changes in rotational discontinuities when such non-ideal effects are introduced \citep{lyu_1989, hau_1991, hau_2016}.  
These theoretical papers used different analytical models based on different simplifying assumptions. Contrary to these papers, we will not use such kind of assumptions. Instead, we will just analyze the observed magnetic hodograms, and show that their shape is incompatible with a gyrotropic pressure.

\section{The magnetopause normal}
When studying the magnetopause with \textit{in situ} measurements, the most basic geometric characteristic to be determined is the normal to its surface (which may vary during the crossing). An accurate determination of the magnetopause normal is actually a fundamental condition for determining reliable estimates of the normal components of both the magnetic and the mass fluxes. Moreover, having a good estimation of the normal direction is also necessary to determine the speed of the structure and its thickness. Quantitatively speaking, to determine the normal component of the magnetic field sufficiently well (assuming that $B_n/|\mathbf{B}|\sim 2\%$), an accuracy of the normal should be of the order of $\delta \theta< 1^o$. In the literature, a good accuracy of determination of the normal is considered to be of the order of 5$\%$ \citep{Denton2018}. 

Beyond determining the normal direction, some "reconstruction methods" can be used to provide a more global view of the large scale structure around the spacecraft. Although these methods have proven to provide remarkable results ~\citep{de_keyser_empirical_nodate, hasegawa_optimal_2005, denton_polynomial_2020} they will not be used here (the first two studies assume the Grad-Shafranov equations to be valid, implying stationary MHD, and are therefore not appropriate to investigate the non-MHD effects such as the FLR effects).

Over the years, several methods have been developed with the purpose to precisely determine the normal direction (see e.g. \citep{haaland2004, shi2019}). The most common is the minimum variance (MVA) introduced with the first measurements of the magnetic field in space \citep{sonnerup1967, sonnerup1998}. This method, which requires single spacecraft measurements, provides a \textit{global} normal, $i.e.$ a single normal vector for each entire time series across the boundary. The tool is based on the assumption that the boundary is a perfectly one-dimensional and stationary layer crossing the spacecraft. Other notable examples are the Generic Residue Analysis (GRA) technique \citep{sonnerup_orientation_2006}, which consist in a generalisation of the MVA to other parameters than \textbf{B}, and the BV method \citep{dorville_rotationalcompressional_2014}, which combines magnetic field and velocity data.
Even though these methods can give an accurate normal determination \citep{dorville_magnetopause_2015}, they provide, like MVA, a \textit{global} normal and thus they cannot provide the necessary basis for investigating the variations of the magnetopause normal within the structure and test the possible departures from mono-dimensionality. Let us finally recall that waves and turbulence, which are always superimposed to the laminar magnetopause profiles, bring strong limitations in the normal direction accuracy for all methods, in particular these global ones.

In this context, multi-spacecraft missions have represented a fundamental step in increasing the accuracy of the normal determination, allowing to determinate the gradients of the measured fields. A notable example is the Minimum Directional Derivative (MDD, \citet{shi_dimensional_2005}) method. This tool generally uses the magnetic field data, but it must be kept in mind that it is not based on specific properties of this field. The MDD technique is a so-called "gradient based method" since the calculation of the normal is based on the experimental estimation of the dyadic tensor $\mathbf{G} = \mathbf{\nabla} \mathbf{B}$. This tensor gradient can be obtained from multispacecraft measurements using the reciprocal vector method ~\citep{chanteur_spatial_1998}. The MDD method consists in diagonalizing the matrix $\textbf{L}=\textbf{G}\cdot \textbf{G}^T$, finding the normal direction as the eigenvector corresponding to the maximum eigenvalue. Moreover, the gradient matrix can also be used for estimating the dimensionality of the boundary from the ratio between the eigenvalues. A way of finding a quantitative determination of this dimensionality was proposed in \citet{Rezeau2017}. 

For the vector $\mathbf{B}$, the MDD method makes use only of the spatial derivatives $\partial_i \mathbf{B}$, which are accessible at each time step thanks to the 4-point measurements today available with multi-spacecraft space missions. In this sense, it is the opposite of the MVA method, which makes use only of the temporal variances of the $\mathbf{B}$ components. It therefore allows for an instantaneous determination of the normal at any point inside the layer, while MVA can only provide a single normal for a full crossing. In addition, contrary to MVA, MDD does not make any assumption about the geometry of the layer (1D variations or not), and about the physical properties of the vector used. Indeed, it can be applied to the magnetic field data but also to any other vector since the property $\nabla\cdot\mathbf{B}=0$ is not used.

However, due to waves and turbulence, the magnetopause can present locally two-dimensional properties that are insignificant for the profiles we are looking for. For this reason, we will focus here on intervals where the magnetopause is mainly one-dimensional, discarding the crossings in which local 2D features are observed. The intervals considered as one-dimensional are those for which $\lambda_{max}\gg\lambda_{int}$. Here $\lambda_{max}$ and $\lambda_{int}$ are defined as the highest and the intermediate eigenvalues of the matrix $\textbf{G}$. In this limit, the ordering between $\lambda_{int}$ and $\lambda_{min}$ ($i.e.$ the smaller eigenvalue) is not relevant in defining the intervals. Specifically, we use the parameter, $D_1 = (\lambda_{max}-\lambda_{int})/\lambda_{max}$, which enables us to quantify this mono-dimensionality of the magnetopause as a function of time.

A more recent tool proposed to study the magnetopause is the hybrid method presented in \citet{Denton2016a, Denton2018}, in which the orientation of the magnetopause is obtained through a combination of the MDD and MVA methods, resulting in an improved accuracy of the normal direction.

The only limitation to the MDD accuracy comes from the uncertainty of the spatial derivatives that it uses. In particular, the local gradient matrix is calculated through the reciprocal vector technique \citep{chanteur_spatial_1998}, which assumes linear variations between the spacecraft. 
Because small-scale waves and turbulence are always superimposed on the magnetopause profiles being searched for, this assumption cannot be well respected without some filtering. This filtering actually leads to introduce part of the temporal information on the variations, but it still allows keeping local information inside the layer whenever one filters only the scales sufficiently smaller than those associated to the full crossing duration. 
The quality of the filtering is therefore the biggest challenge to complete for getting accurate results. For instance, simple gaussian filters done independently on the four spacecraft would provide insufficient accuracy: this can be observed by the fact that, when doing it, the relation $\nabla \cdot \mathbf{B}=0$ is violated in the result. 
In the following section, it is shown how the MDD method can be included in a fitting procedure of the four spacecraft simultaneously and where this relation can be imposed as a constraint. We also show that, when no constraint is added, this procedure justifies the use of MDD with data that are filtered independently. 

\section{A new tool}
The tool we present here, namely GF2 (Gradient matrix Fitting), has been derived from the MDD method. The digit 2 indicates that in the version of the tool that we use here the data are fitted with a 2D model (it can be shown that fitting with a 1D model is mathematically equivalent to the standard MDD technique used with smoothed data). Differently from the original method, we assume that the structure under investigation can be fitted locally ($i.e.$  in each of the small sliding window used along the global crossing), by a two-dimensional model. This does not imply that the magnetopause is assumed globally two-dimensional. As for MDD, the instantaneous gradient matrix $\mathbf{G}$ is obtained from the data using the reciprocal vector's technique ~\citep{chanteur_spatial_1998}. When performing the 2D fit in each sliding window, we then impose some physical constraints, which could be checked only \textit{a posteriori} with the classic MDD method.

The model $\mathbf{G}_{fit}$ is obtained as follows:
\begin{equation}
    \mathbf{G}_{fit}=\mathbf{e}_0\;\mathbf{B}_{e0}'+\mathbf{e}_1\;\mathbf{B}_{e1}'
\end{equation}
where we define $\mathbf{e}_0$ and $\mathbf{e}_1$ as two unit vectors in the plane perpendicular to the direction of invariance and $\mathbf{B}_{e0}'$ and $\mathbf{B}_{e1}'$ as the variation of the magnetic field along these two directions.

\noindent By performing the fit, we impose $\nabla\cdot \mathbf{B}=0$  (as used in MVA but ignored in standard MDD). In the model, this can be written as:
    \begin{equation}
        \mathbf{e}_0\cdot\mathbf{B}_{e0}'+\mathbf{e}_1\cdot\mathbf{B}_{e1}'=0
        \label{eq:BC}
    \end{equation}

In order to fit the experimental $\mathbf{G}$ by the model $\mathbf{G}_{fit}$, the following quantity has to be minimised
\begin{equation}
\begin{split}
    D_2&=\Tr [(\mathbf{G}_{fit}-\mathbf{G}).(\mathbf{G}_{fit}-\mathbf{G})^T]\\
    &=\mathbf{B}_{e0}'^2-2\mathbf{e}_0.\mathbf{G}.\mathbf{B}_{e0}'+\mathbf{B}_{e1}'^2-2\mathbf{e}_1.\mathbf{G}.\mathbf{B}_{e1}'+Tr(\mathbf{G} \mathbf{G}^T)
\end{split} 
\end{equation}
We can disregard the last term, since it is independent of the fit parameters. To impose the physical constraints, we use Lagrange multipliers, minimizing:
\begin{equation}
\begin{split}
    D_2=&\mathbf{B}_{e0}'^2-2\mathbf{e}_0.\mathbf{G}.\mathbf{B}_{e0}'+\mathbf{B}_{e1}'^2-2\mathbf{e}_1.\mathbf{G}.\mathbf{B}_{e1}'+2\lambda(\mathbf{e}_0\cdot\mathbf{B}_{e0}'+\mathbf{e}_1\cdot\mathbf{B}_{e1}')\\
    =&\mathbf{B}_{e0}'^2-2\mathbf{e}_0.(\mathbf{G}-\lambda\mathbf{I}).\mathbf{B}_{e0}'+\mathbf{B}_{e1}'^2-2\mathbf{e}_1.(\mathbf{G}-\lambda\mathbf{I}).\mathbf{B}_{e1}'
\end{split} 
\label{eq:LMulti}
\end{equation}
By assuming in the first approximation that the direction of invariance $\mathbf{e}_2$ is known, we can choose the two vectors $\mathbf{e}_0$ and $\mathbf{e}_1$ as an arbitrary orthonormal basis for the plane of variance. For performing the minimisation, we have just to impose equal to zero the derivatives with respect to $\mathbf{B}_{e0}'$, $\mathbf{B}_{e1}'$ and $\lambda$, obtaining Equation \ref{eq:BC} and:
\begin{equation}
    \mathbf{B}_{e0}'=\mathbf{e}_0.(\mathbf{G}-\lambda\mathbf{I})
\end{equation}
\begin{equation}
    \mathbf{B}_{e1}'=\mathbf{e}_1.(\mathbf{G}-\lambda\mathbf{I})
\end{equation}
By introducing these two equations in Equation \ref{eq:BC} we obtain:
\begin{equation}
    \lambda=\frac{G_{00}+G_{11}}{2},
    \label{constraint}
\end{equation}
from which we get the values of $\mathbf{B}_{e0}'$ and $\mathbf{B}_{e1}'$. At this point, the matrix $\mathbf{G}_{fit}$ is fully determined. We can then look for its eigenvalues and eigenvectors, as in the standard MDD method, and get the normal $\mathbf{n}$ and the tangential directions $\mathbf{t}_1$ ($i.e.$ the one orthogonal to the direction of invariance) from this smooth fit.

The choice of the direction of invariance has actually no major influence on the determination of the normal direction, neither on the estimation of the 2D effects. For large 2D effects, one could choose the direction of minimum variance obtained by applying directly the standard MDD method to the data. Nevertheless, for almost 1D cases (the most common situation), the spatial derivatives in the tangential directions are generally much smaller than the noise, so this result is not reliable. We simply choose here the constant M direction given by MVA, which is often considered as the direction of the X line if interpreted in the context of 2D models of magnetic reconnection (\textit{cf.} for instance \citet{phan_electron_2013} for a typical use of this choice and \citet{aunai_orientation_2016, liu_collisionless_2018,Denton2018} for discussions about it).

Finally, another useful by-product of the method can be obtained: comparing the spatial derivatives and the temporal ones and using a new fitting procedure, we can compute the two components of the velocity of the structure $V_{n0}$ and $V_{t1}$ with respect to the spacecraft. Only the motion along the invariant direction then remains unknown.

\subsection{Normal from ions mass flux}
This tool can be easily adapted to any other vector dataset by just changing the physical constraint. In particular, we chose to study the structure using the ion mass flux data. In this case we impose mass conservation $\nabla\cdot \boldsymbol{\Gamma}_i=-\partial_t n_i$ (with $\boldsymbol{\Gamma}_i=n_i\mathbf{u}_i$). Eq.\ref{eq:BC} now writes
\begin{equation}
    \mathbf{e}_0\cdot\boldsymbol{\Gamma}_{e0}'+\mathbf{e}_1\cdot\boldsymbol{\Gamma}_{e1}'+\partial_t n_i=0
    \label{eq:BC1}
\end{equation}
Therefore, when using the Lagrange multipliers, Eq.\ref{eq:LMulti} changes to:\begin{equation}
\begin{split}
    D_2=&\boldsymbol{\Gamma}_{e0}'^2-2\mathbf{e}_0.\mathbf{G}.\boldsymbol{\Gamma}_{e0}'+\boldsymbol{\Gamma}_{e1}'^2-2\mathbf{e}_1.\mathbf{G}.\boldsymbol{\Gamma}_{e1}'+2\lambda(\mathbf{e}_0\cdot\boldsymbol{\Gamma}_{e0}'+\mathbf{e}_1\cdot\boldsymbol{\Gamma}_{e1}'+\partial_t n_i)\\
    =&\boldsymbol{\Gamma}_{e0}'^2-2\mathbf{e}_0.(\mathbf{G}-\lambda\mathbf{I}).\boldsymbol{\Gamma}_{e0}'+\boldsymbol{\Gamma}_{e1}'^2-2\mathbf{e}_1.(\mathbf{G}-\lambda\mathbf{I}).\boldsymbol{\Gamma}_{e1}'+2\lambda\partial_t n_i
\end{split} 
\end{equation}
By using the same algorithm as above, the constraint can now be written as: 
\begin{equation}
    \lambda=\frac{G_{00}+G_{11}+\partial_t n_i}{2}
\end{equation}
\subsection{Dimensionality index} \label{sec:dim}
From this procedure, we can also derive another significant result: we can obtain an indicator of the importance of the 2D effects in the profiles, free of the parasitic noise effects. Specifically, we can estimate the variation of the magnetic field along the normal by projecting the $\mathbf{G}_{fit}$ matrix along it $\text{var}_n=|\mathbf{G}_{fit}. \mathbf{n}|$. Consequently, if we designate the variation along $\textbf t_1$ as $\text{var}_t$, we can introduce a new dimensionality index:
\begin{equation}
    \mathscr{D}_{GF2}=\frac{\text{var}_n-\text{var}_t}{\text{var}_n}
    \label{eq:dim}
\end{equation}
This index can usefully be compared with the instantaneous index $D_1 = (\lambda_{max}-\lambda_{int})/\lambda_{max}$ of \citet{Rezeau2017}.

\section{Expected accuracy and tests of the tool}
\begin{figure}
\centering
\includegraphics[width=0.9\columnwidth]{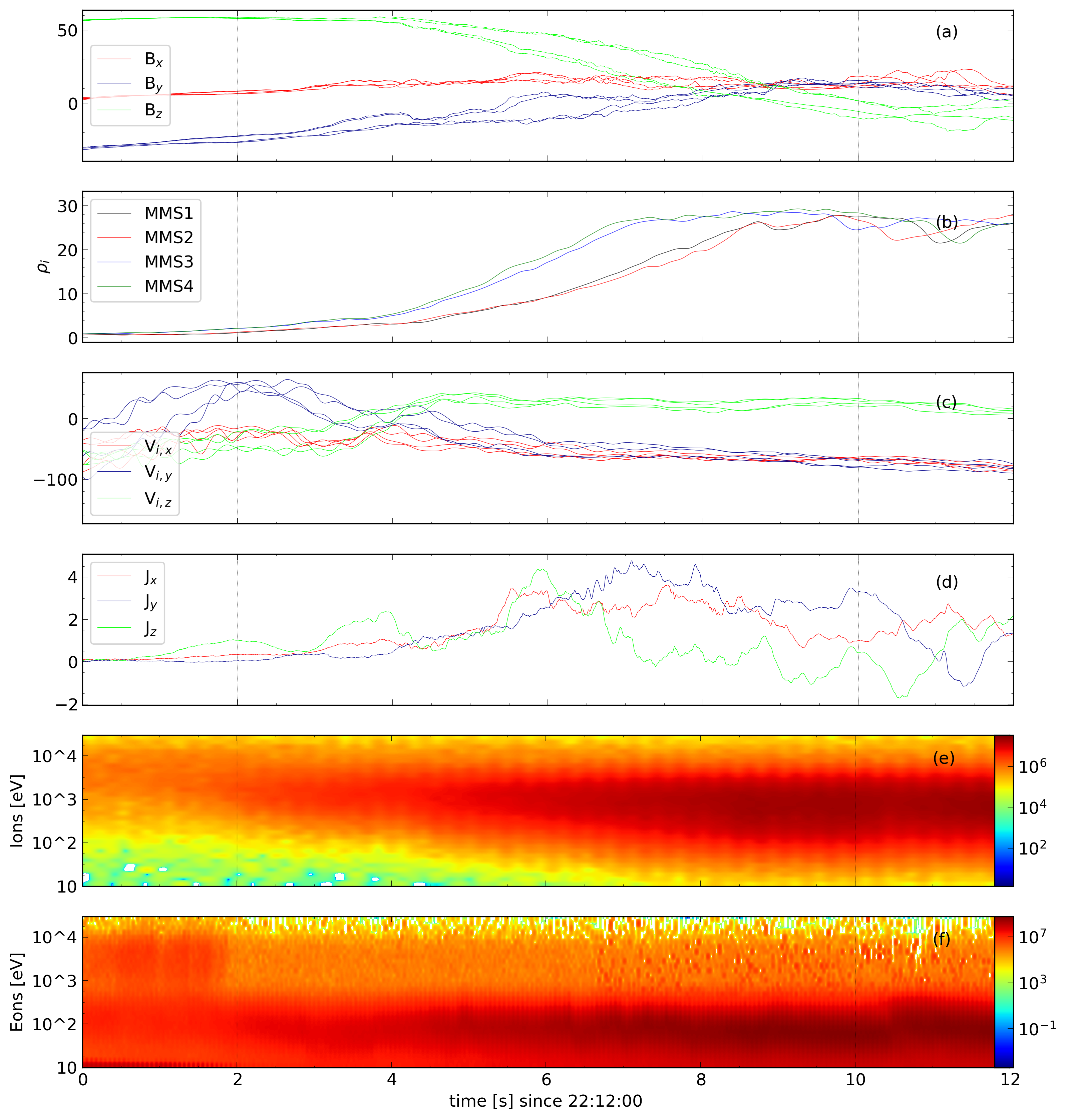}
\caption{Main features of the crossing of the 28th December 2015. From top to bottom: ($a$) the magnetic field (in $nT$), ($b$) the ion particle density (in $m^{-3})$, ($c$) ion velocity (in $km/s$), ($d$) total current (computed from the curlometer \citep{dunlop_analysis_1988}, in $nA/m^{2}$), ($e$) the ion and ($f$) electron spectrograms (energies are shown in $eV$). Vertical lines indicate the time interval chosen for the case study.}
\label{fig:event2}
\end{figure}

In this section we test the accuracy of the GF2 tool. To accomplish this, we exploit a case crossing, which will be investigated in detail in the following section. The crossing considered comes from MMS data~\citep{burch_phan_2016}, taking place at around 22:11 on 28th December 2015. For this study we use data from the FluxGate Magnetometers (FGM, \citet{russell_magnetospheric_2016}), providing the magnetic field data, the Electric Double Probe (EDP, \citet{lindqvist_spin-plane_2016, ergun_axial_2016}), for the electric field, and Dual Ions and Electrons Spectrometer instrument (DIS, DES, \citet{pollock_fast_2016}), for plasma measurements. An overview of the event is shown in Fig. \ref{fig:event2}, where both the magnetic ﬁeld and ion bulk velocity are given in Geocentric Solar Ecliptic (GSE) coordinates. For this event, the spacecraft are located in [7.6, -6.7, -0.8] $R_E$ in GSE coordinates (where $R_E$ is the Earth's radius).

The temporal interval in which we observe the shear in the magnetic field and the crossing in the particle structure is about 8 seconds, enough to allow for high resolution for both sets of measurements. The crossing is chosen by also analyzing the dimensionality of the magnetic field measurements averaged along the crossing. In particular, the dimensionality parameter defined in Eq. \ref{eq:dim}, denoted as $\mathscr{D}_{GF2}$, and the one introduced in \citet{Rezeau2017}, denoted as $D_1$, were considered. In this interval, indeed, we have $D_{1, mean}=0.97 \pm 0.03$ while $\mathscr{D}_{GF2}=0.89\pm0.06$, both highlighting that the crossing exhibits one-dimensional features throughout the time interval.
We remind here that in burst mode, the frequency of magnetic field measurements is 132Hz while it is 6,67 Hz for ions. To conduct the following study, it is necessary to interpolate all measurements at the same times. We did it by testing two sampling frequencies: the magnetic field and the ion ones. The results obtained are consistent with the two methods. All figures shown in the paper are obtained with the sampling times of the MMS1 magnetic field. Furthermore, the crossing is observed quasi-simultaneously for the two quantities, with a large interval where the two kinds of results can be compared. 
\begin{figure}
    \centering
    \includegraphics[width=\textwidth]{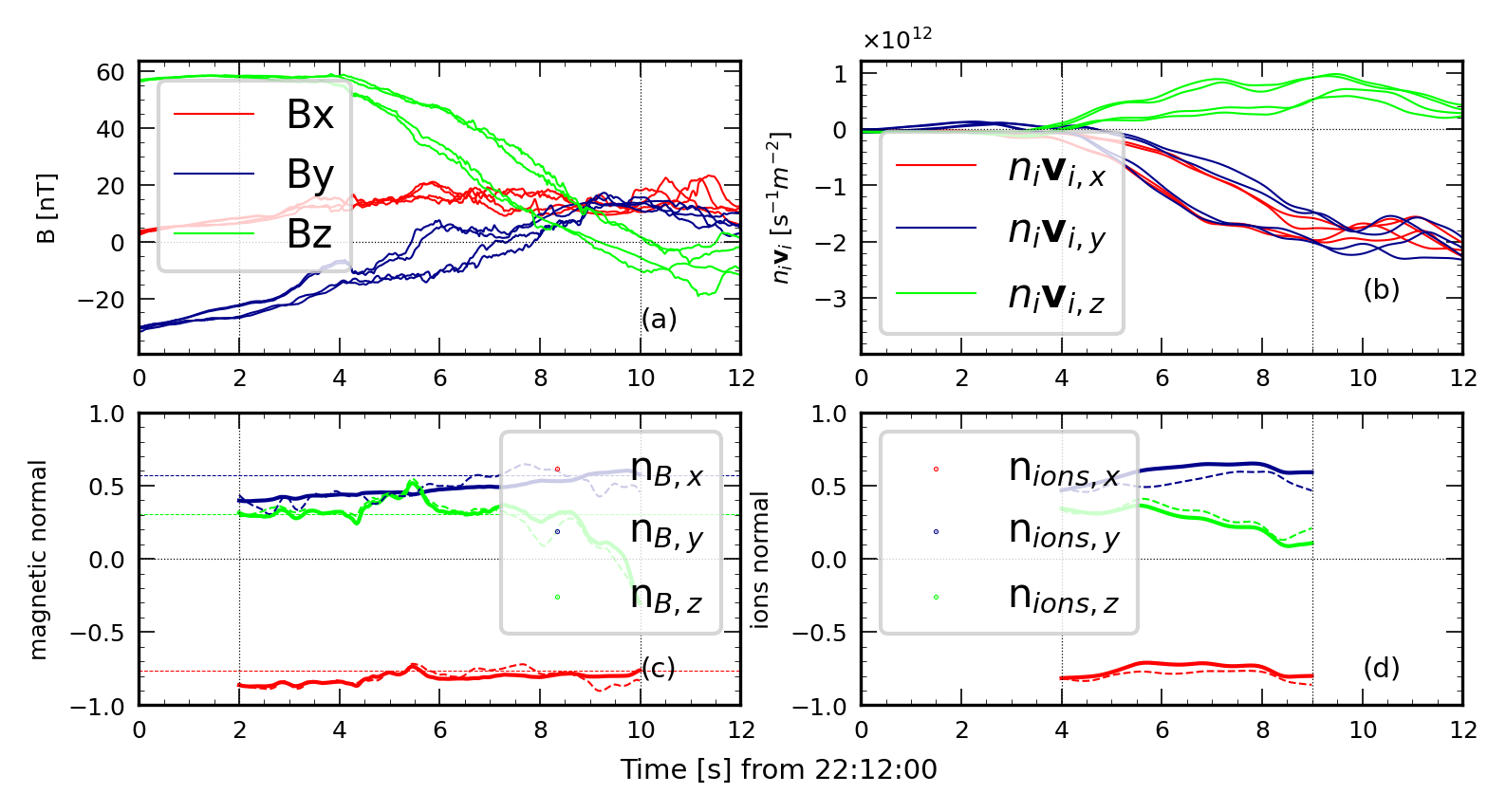}
    \caption{Comparison for the normals obtained with GF2 with respect to the MDD tools. $(a)$ shows the magnetic field and $(b)$ the ion mass flux, measured by the four MMS spacecraft. $(c)$ and $(d)$ shows the magnetic and the ion normal, respectively. The continuous (resp. dashed) line correspond to the components of GF2 (resp. MDD) normal. Horizontal dotted lines indicate the MVA normal obtained along the whole interval. Vertical dashed lines correspond to the time interval boundaries for the crossing, which are different for the magnetic field and the ion mass flux.}
    \label{fig:comparison}
\end{figure}

As a first test, we compare in Fig.\ref{fig:comparison} the normals obtained by GF2 and those by the standard MDD technique (using data smoothed in a 0.31s time-window), for both the magnetic field and the ion data. For reference, we also compare the result of $GF2_B$ with the MVA one.

Vertical dashed lines indicate the time interval during which all the satellites are inside the boundary. We observe that the time required for the ions flux to complete the crossing (of about 5$s$) is shorter than for the magnetic field (about 8$s$). To perform a quantitative analysis of the differences, we studied the angles between the different normals obtained through GF2, MDD and MVA, as shown in Fig. \ref{fig:angles}.a.

\begin{figure}
    \centering
    \includegraphics[width=\columnwidth]{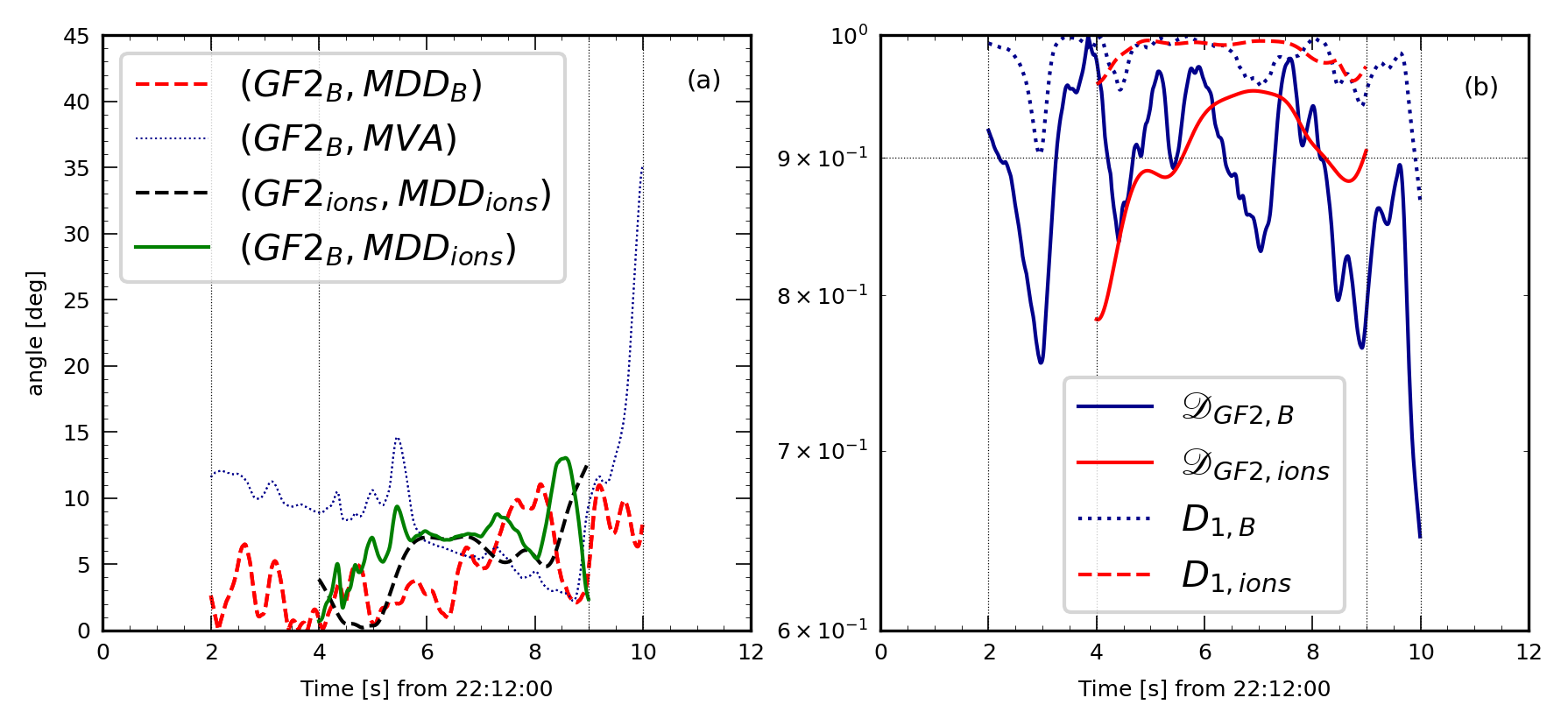}
    \caption{$(a)$ Angle between the normals obtained using the state-of-the-art tools (MDD, MVA) and GF2. The subscript $B$ and $ions$ indicates whenever the magnetic field or the ion flux measurements are used. $(b)$ Dimensionality of the structure as a function of time; here both the $\mathscr{D}_{GF2}$ (continuous line) and the $D_1$ (dashed line) indices are shown, for both the magnetic field (blue) and ions (red) data.}
    \label{fig:angles}
\end{figure}

The first striking result is that all these results are quite consistent. Almost all the directions are less than ten degrees apart from each other, with an average difference of about five degrees. The major exception concerns the comparison between MVA and $GF2_B$ during the last second of the interval where the two directions appear to be up to 35 degrees apart. This can be explained by the fact that the local normals are observed (by $GF2_B$ as well as by $MDD_B$)  to differ noticeably in this part from their averaged value and that MVA is not able to detect such a change. Looking in more detail, we can see a slight difference between the first part of the crossing (between 2$s$ and  $\sim$6.5$s$), where the two normals $GF2_B$ and $MDD_B$ differ by less than 5 degrees, and the second part, where the angle between the normals can be up to ten degrees (probably due to a smaller ratio signal/noise for the gradient matrix $\textbf{G}$). The normals derived from ion measurements are not much different from those derived from the magnetic field, showing that the particle and magnetic structures are approximately identical. In Fig. (\ref{fig:angles}.b), the dimensionality of the structures is analyzed as a function of time, by using both the $\mathscr{D}_{GF2}$ and the $D_1$ (\citet{Rezeau2017}) parameters, as explained above. Even if the numerical values of the two indices are slightly different, they both indicate structures close to one-dimensionality in the first part, with a -small but significant- decrease in the second part. This increased departure from mono-dimensionality can explain the slight difference between the two parts when comparing the normals from standard MDD and GF2 techniques.

The present test does not allow us to state that GF2 is more accurate than standard MDD (this will be checked in future work by comparing the two tools in a global simulation involving realistic turbulence) but it shows at least a good agreement between the two approaches. We will see in the following that this accuracy is anyway sufficient to prove the role of FLRs.

In order to smooth the small fluctuations over the time interval and to reduce the statistical error associated with the determination of the normal, we can compare the directions averaged along the crossing time. Mean values obtained through the tools presented above are shown in Table \ref{tab:normals}.
\begin{table}
    \centering
    \begin{tabular}{ccc}
       \textbf{Model}  & \textbf{Normal} [GSE] & angle with\\
         & & $\mathbf{n}_{GF2, B}$ [deg]\\
       \hline
       $\mathbf{n}_{GF2, B}$  &  [0.82, -0.49, -0.29]& x  \\
       $\mathbf{n}_{GF2, ions}$  & [0.76,  -0.59,  -0.26]&  7.2 \\
       MVA  & [0.76, -0.57, -0.30]&  6.1 \\
       MDD  & [0.83, -0.49, -0.28]& 0.7\\
       Denton  & [0.82, -0.49,-0.29]&  0.4 \\
       \\
       \end{tabular}
    \caption{Magnetopause normal vectors obtained with the main tools presented above averaged in the time interval and their angle with respect to the normal obtained with GF2 using the magnetic field data (in degrees).}
    \label{tab:normals}
\end{table}
Here we observe that all the averaged normals differ by less than 10 degrees. Specifically, we observe that the normals obtained with GF2, MDD and \citep{Denton2018} are similar, with a difference of less than one degree (with ours being closer to the one from \citet{Denton2018}). MVA normal, instead, differs around 6 degrees from all these other normals. Finally, we also observe that the one computed with ions flux data is the most distant. This is interpreted to be due to the higher uncertainty of particles measurements.

\section{Case study}\label{sec:role}
In this section, we undertake a detailed analysis of the previously mentioned crossing case by employing the normal obtained using the GF2 tool. Here, we focus on the time interval between 2~$s$ and 9~$s$ in Fig \ref{fig:event2}. To mitigate the potential influence of non-unidimensionality effects, we chose to exclude the last second of the time interval studied in the preceding section for the magnetic field (where both $\mathscr{D}_{GF2}$ and $D_1$ show that the structure is less one-dimensional and where we observe that the normal is more different from the averaged one). To carry out this analysis, we study the hodogram of the magnetic field in the tangential plane. Here the tangential results are presented in a basis ($\mathbf{T}_1, \mathbf{T}_2$) chosen as: 
\begin{equation}
    \mathbf{ T}_1=\mathbf{n}_{mean} \times \mathbf{\hat b}    
\end{equation}
\begin{equation}
    \mathbf{ T}_2=\mathbf{n}_{mean} \times \mathbf{T}_1
\end{equation}
where $\mathbf{\hat b}=\mathbf{B}/|\mathbf{B}|$ and $\mathbf{n}_{mean}$ is the directions of the averaged normal in the chosen time interval. Note that the choice of the reference frame ($\mathbf{T}_1, \mathbf{T}_2$) is just a convention. The shape of the hodogram remains unaffected by this choice except for the corresponding rotation in this tangential plane. The direction $\mathbf{t}_1$, which characterizes the direction of the second dimension of the model in GF2 and which is also in the tangential plane is generally different from $\mathbf{T}_1$.
 
\begin{figure}
\centering
\includegraphics[width=0.8\columnwidth]{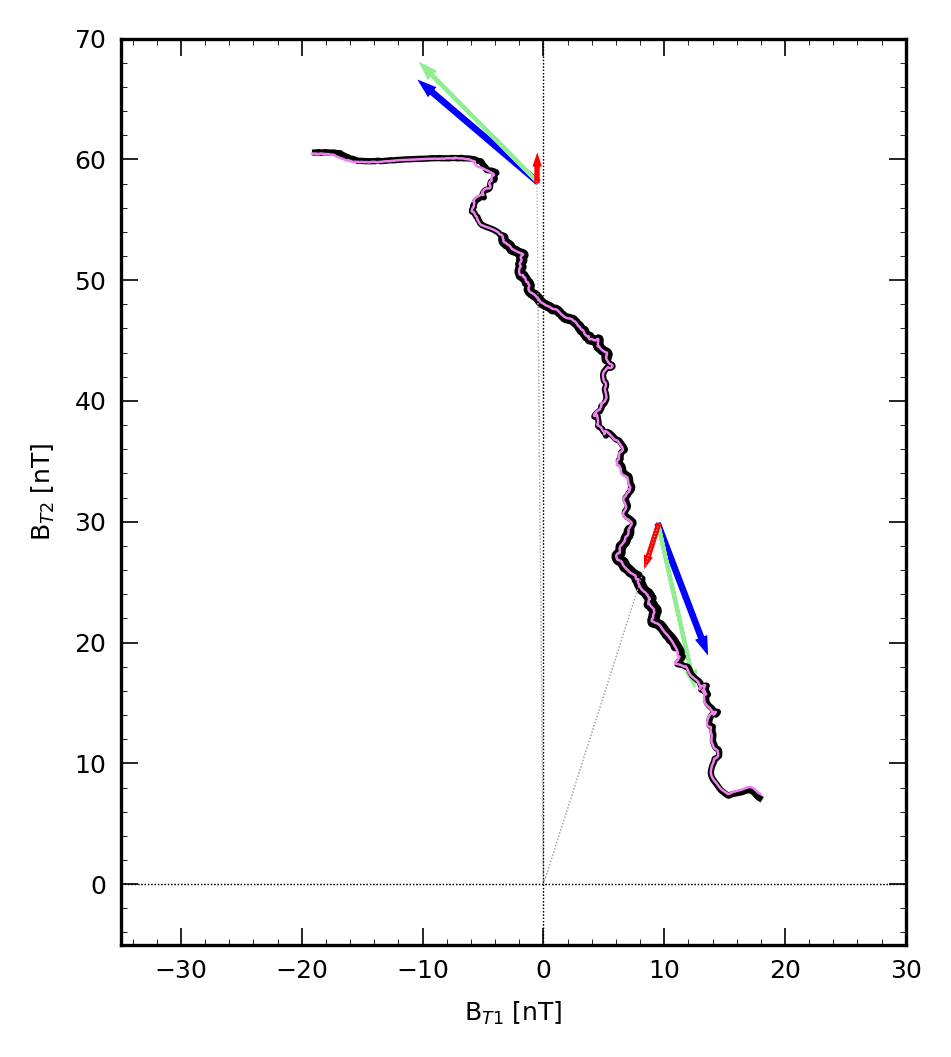}
\caption{Hodogram in the tangential plane of the magnetic field for a magnetopause crossing by MMS in 28.12.2015 from 22:12:02 to 22:12:09. See text for the significance of the arrows. $B_{T1}$ and $B_{T2}$ are the projections of \textbf{B} along the tangential directions computed as described in the text. The black line (resp. violet) is the hodogram when the $\mathbf{n}_{mean}$ (resp. $\mathbf{n}$) value is used to define the reference frame.}
\label{fig:hodogram}
\end{figure}

If CTD was valid everywhere, the hodogram of the magnetic field in the tangential plane for a rotational discontinuity would correspond to a circular arc with constant radius while a shock would correspond to a radial line (as shown in Fig. \ref{fig:cartoon}). For this reason, the hodogram is a good tool to recognize the cases for which the CTD fails at describing the magnetopause. The hodogram for this case is shown in Fig. \ref{fig:hodogram}. We observe a clear "linear" (although not radial) hodogram. This non-radial variation of the magnetic field although not predicted by CTD, is a striking feature of the hodogram. It cannot be explained by a departure from the 1D assumption since we have measured that the crossing can be considered as one-dimensional to a good degree of accuracy. It is therefore due to an intrinsic property of the layer itself. Also, in Fig. \ref{fig:hodogram}, we present the hodogram derived from the local normal (un-averaged, violet line). It is clear that averaging does not affect the shape of the hodogram.

To further analyze the causes of the disagreement between the hodogram of this case crossing and what is expected from CTD, we compare the different terms of the tangential momentum equation and Faraday/Ohm's law. As discussed above, indeed, these two equations are responsible for the distinction between the rotational and tangential discontinuities in CTD. This is the object of Figure \ref{fig:Ohm_momentum1}, where we plot the different terms of the two equations projected along the $\mathbf{n}_{mean}$ and $\mathbf{t}_{1,mean}$ directions obtained using the GF2 tool (averaged over the whole time interval). The influence of the averaging of the $\mathbf{t}_{1,mean}$ direction on the results is discussed in Appendix \ref{appendix}. We do not show the quantities along the direction of invariance, which are dominated by noise. The current and the gradient matrix for the pressure term are obtained via the reciprocal vector method described in \citet{chanteur_spatial_1998}.
\begin{figure}
\centering
\includegraphics[width=\textwidth]{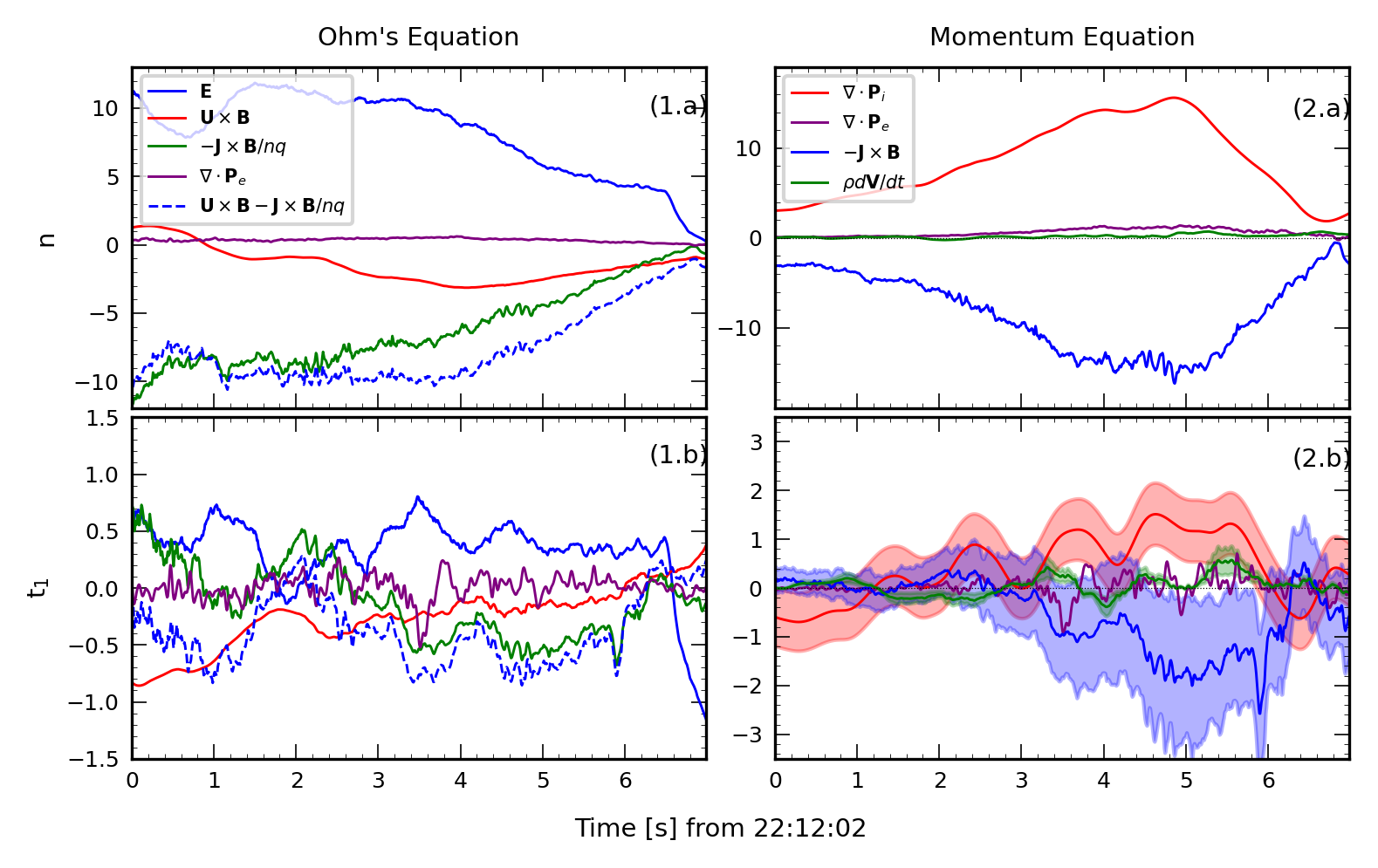}
\caption{\label{fig:Ohm_momentum1} Terms of the Ohm's law (panel 1, units of m$V/m$) and the momentum equation (panel 2, units of $10^{-15}kg\,m/s^2$), projected in the normal direction \textit{n} ($a$) and in the tangential direction $t_1$ ($b$). To reduce the noise, a running average with a time window of 0.35s is applied to the electric field measurements. Shaded regions in panel $2.b$ represent the estimated uncertainties of the divergence of the pressure (red), the $\mathbf{J}\times\mathbf{B}$ (blue) and the classic inertial term (green). Concerning the Ohm's law, we included the sum $\mathbf{U}\times\mathbf{B}-\mathbf{J}\times\mathbf{B}/nq$ to facilitate the readability (blue dashed line).}\textit{N.b.} The terms of the tangential Faraday/ Ohm's law used in the text are just the derivatives of the ones in ($a$) (apart from a $\pi/2$ rotation).
\end{figure}

Concerning the Ohm's law (Figure \ref{fig:Ohm_momentum1}, panel 1), we see that the electric field is well counter-balanced by the $\mathbf{u}\times\mathbf{B}$ and $\mathbf{J}\times\mathbf{B}/nq$ terms (ideal and Hall terms). Outside the layer, on both sides, the ideal Ohm's law is satisfied, as assumed in CTD (this is not visible on the figure, which is a zoom on the inner part of the layer, and where the Hall term is important). It has been shown in the literature that $\nabla \cdot \mathbf{P}_e$ is not always negligible in the Ohm's law and that it can even be dominant close to an Electron Diffusion Regions (EDR). This has been predicted theoretically \citep{hesse_diffusion_2011, hesse_electron_2014} and observed experimentally \citep{torbert_estimates_2016, genestreti_how_2018}, but it is not the case for events like this one. We observe that at approximately $3.5$ seconds, the $\nabla \cdot \textbf{P}_e$ is not entirely negligible along the tangential direction (a similar peak can also be observed in panel $2.b$ for the term associated with the electron pressure in the momentum equation). However, during this time interval, this value is not dominant, this term being smaller than both the electric field and the $\mathbf{J}\times\mathbf{B}/nq$ components. Furthermore, this effect exhibits a local characteristic, as $\nabla \cdot \textbf{P}_e$ is only non-negligible within a small sub-interval (with respect to the magnetopause temporal width). It is therefore not likely to be indicative of proximity to a reconnection point.

Concerning the momentum equation, shown in panel 2 of Figure \ref{fig:Ohm_momentum1}, we observe that, in the normal direction, the $\mathbf{J}\times\mathbf{B}$ term is counter-balanced by the divergence of the ion pressure tensor, as expected. But, if the isotropic condition assumed in CTD was valid, we would expect the divergence of the ion pressure tensor to be zero in the tangential direction, or at least negligible with respect to the inertial term $\rho d\mathbf{u}/dt$. On the contrary, we observe that the $\mathbf{J}\times\mathbf{B}$ term along $\mathbf{t}_1$ is of the same order of magnitude as the divergence of the ion pressure tensor, and one order of magnitude larger than all the other terms. Panel $2.b$ also shows an estimation of the error on the relevant terms: $\mathbf{J}\times\mathbf{B}$, $\nabla \cdot \mathbf{P}_i$ and the classical inertial term. It is known that measurements errors are difficult to estimate, especially at small scales. In order to validate our results, however, we sought to obtain an upper bound of the error associated with the quantities of interest. For that purpose, an overestimation of the uncertainty of the measurements (acquired as the maximum during the crossing of the errors available in FPI datasets for the pressure tensor and from the FGM nominal error for the magnetic field) was exploited. These values are propagated as a statistical ($i.e.$ quadratic) error (by assuming that the errors on the reciprocal vectors can be neglected with respect to that of other physical quantities).

From panel $2.b$ of Fig. \ref{fig:Ohm_momentum1}, we see that the $\mathbf{J}\times\mathbf{B}$ and the $\nabla \cdot \mathbf{P}_i$ terms are pointing in opposite directions and balancing each other. If valid in the first part of the interval, this conclusion cannot be safely trusted due to measurement uncertainty, but we observe that in the middle part (particularly between 3.5$s$ and 6$s$) it is evident that the two quantities counterbalance each other while the classical inertia term $\rho d\mathbf{u}/dt$ is much lower with respect to the others. This proves that the tangent $\nabla \cdot \mathbf{P}_i$ term plays a fundamental role in the magnetopause equilibrium. 

This point can be emphasized also by analyzing the hodogram. In Fig. \ref{fig:hodogram}, the arrows are directed along the directions of the tangential plane that are physically relevant for the problem: $i)$ the tangent to the hodogram (green), which indicates the total variation of $\mathbf B_t$; $ii)$ the radial direction (red), which corresponds to the plasma compression; $iii)$ the $\nabla \cdot \mathbf{P}_{it}$ direction (blue), which is the direction of the divergence of the ion pressure tensor in the tangential plane, and therefore corresponds to a term which is absent in CTD. The relative lengths of the arrows are chosen proportional to the corresponding term magnitudes. These directions are averaged in two sub-intervals (bold hodogram). The striking result is that the total variation is mainly determined by the non-classic term $\nabla \cdot \mathbf{P}_{it}$ and not by the radial classic one. This explains the very recurrent (even if not reported in the literature hitherto) feature that the hodograms are almost linear but not radial.

\subsection{Comparison of the width of the magnetopause to relevant physical lengths}
\begin{figure}
    \centering
    \includegraphics[width=0.8\columnwidth]{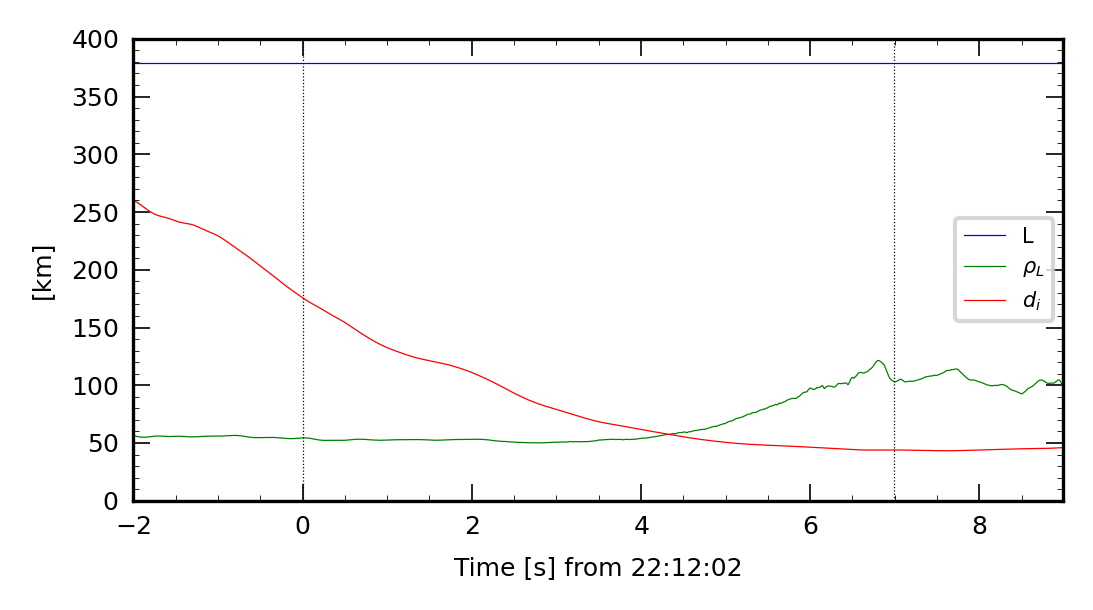}
    \caption{Comparison of the magnetopause width ($L$) with the ion inertial length ($d_i$) and the ions Larmor radius ($\rho_L$). Vertical lines highlight the considered temporal interval. }
    \label{fig:scales}
\end{figure}
Finally, we compare the width of the magnetopause ($L$) to the two main ion-related lengths: the ion Larmor radius ($\rho_L$) and the ion inertial length ($d_i$). The magnetopause width is estimated using the normal velocity obtained from the $GF2$ tool. By averaging the velocity of the magnetopause in the normal direction, we can estimate $L=\mathbf{V}_{n, mean}\Delta t$ (where $\Delta t$ is the time length of the full crossing). These three scales are shown in Fig.\ref{fig:scales}.
We observe that this width is larger than the ion Larmor radius and the ion inertial length all across the crossing, but only two to five times larger, which appears sufficient to drive the observed kinetic effects.

\section{Ion pressure tensor analysis}
\begin{figure}
    \centering
    \includegraphics[width=\columnwidth]{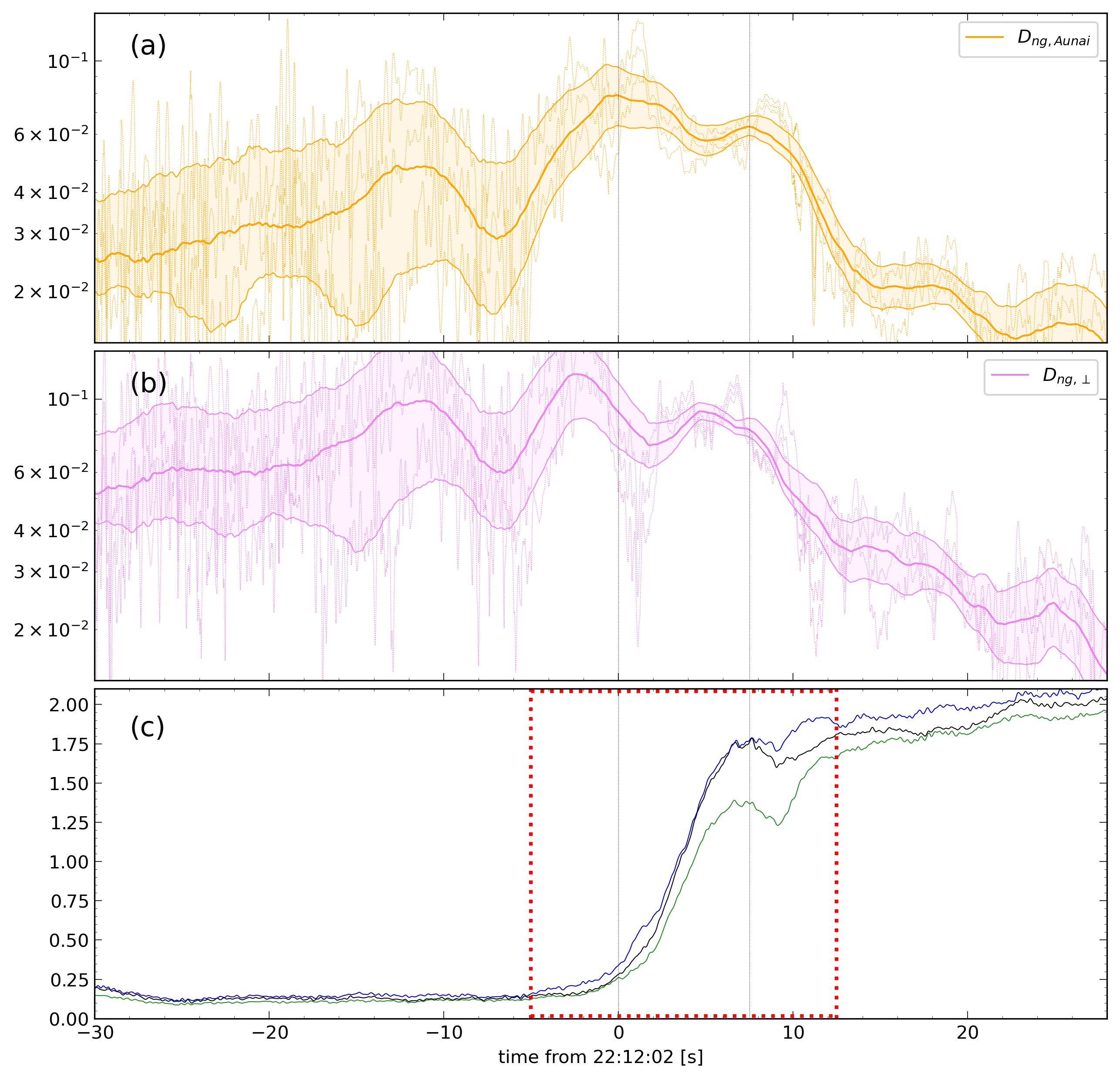}
    \caption{Panels $a$ and $b$ show the evolution of the $D_{ng, \perp}$  and $D_{ng, Aunai}$ \citep{aunai2013} indices, respectively, along with their estimated uncertainties. Thin lines correspond to the real-time values while thick lines to an averaged window of 1 s; ($c$) Evolution of the eigenvalues of the $\mathbf{P}_i$ matrix (averaged on the four spacecraft). The dotted line indicates the magnetopause crossing. The red dotted lines in panel $c$ highlight the time interval studied in Fig. \ref{fig:VDFs}}
    \label{fig:non-gyrotropy}
\end{figure}
To further investigate the question of the ion non-gyrotropy with respect to the magnetic field and quantify this effect, let us now examine the properties of the ion pressure tensor and introduce a new non-gyrotropy index. For that purpose, we define the matrices $\mathbf{P}_\parallel=p_\parallel \mathbf{bb}$, where $\mathbf{b}=\mathbf{B}/|\mathbf{B}|$ and $p_\parallel=\mathbf{b} \cdot \mathbf{P}_i \cdot\mathbf{b}$, and $\mathbf{P}_\perp=\mathbf{P}_i-\mathbf{P}_\parallel$. By defining $p_1$ and $p_2$ the maximum and intermediate eigenvalues of the $\mathbf{P}_\perp$ matrix, we define:
\begin{equation}
    D_{ng, \perp}=\frac{p_1-p_2}{p_1 + p_2}
\end{equation}
In Fig. \ref{fig:non-gyrotropy}.a, this parameter is compared to the non-gyrotropy index presented in \citet{aunai2013}. The two indices define nongyrotropy differently,  \citep{aunai2013}'s index defining nongyrotropy as the ratio of the nongyrotropic to the gyrotropic part of the tensor (instantaneous), while ours makes use of the 2D modeling of the data used in GF2 (averages on sliding windows). We note how both indices are significantly different from zero, approximately of the order of 0.1 within the boundary, corresponding to clearly present, although not predominant, non-gyrotropic effects.
We note a decrease in both indices outside the magnetopause, as expected, but it is worth noting also that, despite a continuous decrease,  these indices remain relatively high in the time interval just preceding the crossing, in a region where magnetic field, density and pressure tensor are almost constant. This can be understood by noting that an ion velocity gradient is observed in this interval, suggesting that the non-diagonal terms of the pressure tensor could be due there to a kind of gyroviscous effect, the non-diagonal terms of the pressure tensor \citep{braginskii_transport_1965} being due to FLRs \citep{stasiewicz_finite_1993}. One must take care that, in this interval, the pressure tensor has low values characterized by larger relative errors, which could partially influence this result. 
To further analyze this question, we have estimated the uncertainties on both non-gyrotropy indices. This estimation is derived from the nominal uncertainties of the FPI dataset. The diagonal terms have higher values and lower relative errors. Concerning the time interval before the crossing that we discuss here, the diagonal terms have errors of approximately 5\%, whereas off-diagonal terms have an average relative error about 50\% . We observe on Fig. \ref{fig:Ohm_momentum1} that this way of estimating the uncertainty well encompasses the variance of the results. It confirms that, within the crossing interval, all relative errors are smaller than 10 \%, as considered in the Ohm's law study (Fig. \ref{fig:Ohm_momentum1}).\\
In addition, a preliminary study appears to confirm the validity of the gyroviscous interpretation. Using the theoretical expressions given in \citet{Stasiewicz1989}, we can compare the variations of the non-diagonal terms of the pressure tensor with the spatial derivatives of the flow velocity, and evidence a fairly good correlation (see Appendix \ref{app:gyroviscous}).\\
Fig. \ref{fig:non-gyrotropy} (panel $c$) also shows the evolution of the eigenvalues of the $\mathbf{P}_i$ tensor, averaged on the four spacecraft. This figure shows how outside of the magnetopause the three eigenvalues tend to converge towards each other meaning that these media are close to isotropy. However, inside the magnetopause, we note a transition in the behavior of the intermediate eigenvalue, shifting from a value close to the minor one to being closer to the major eigenvalue. The minor eigenvalue exhibits a significant deviation from the other two towards the last part of the crossing.

Focusing on the temporal interval marked by the red square in Fig. \ref{fig:non-gyrotropy}, this transition is further investigated in Fig. \ref{fig:VDFs} where we show the ions' distribution functions in the tangential plane (with respect to the magnetopause) for four different intervals during the crossing, highlighting the non-gyrotropy of the ions' distribution function over time. VDFs (printed using a linear 2D interpolation on a cartesian grid in the chosen plane using the Pyspedas library) are here averaged in the corresponding time intervals framed with the same color as in the bottom plot where the eigenvalues of the ion pressure tensor are plotted again (the time length decreases as the density increases).

\begin{figure}
    \centering
    \includegraphics[width=\columnwidth]{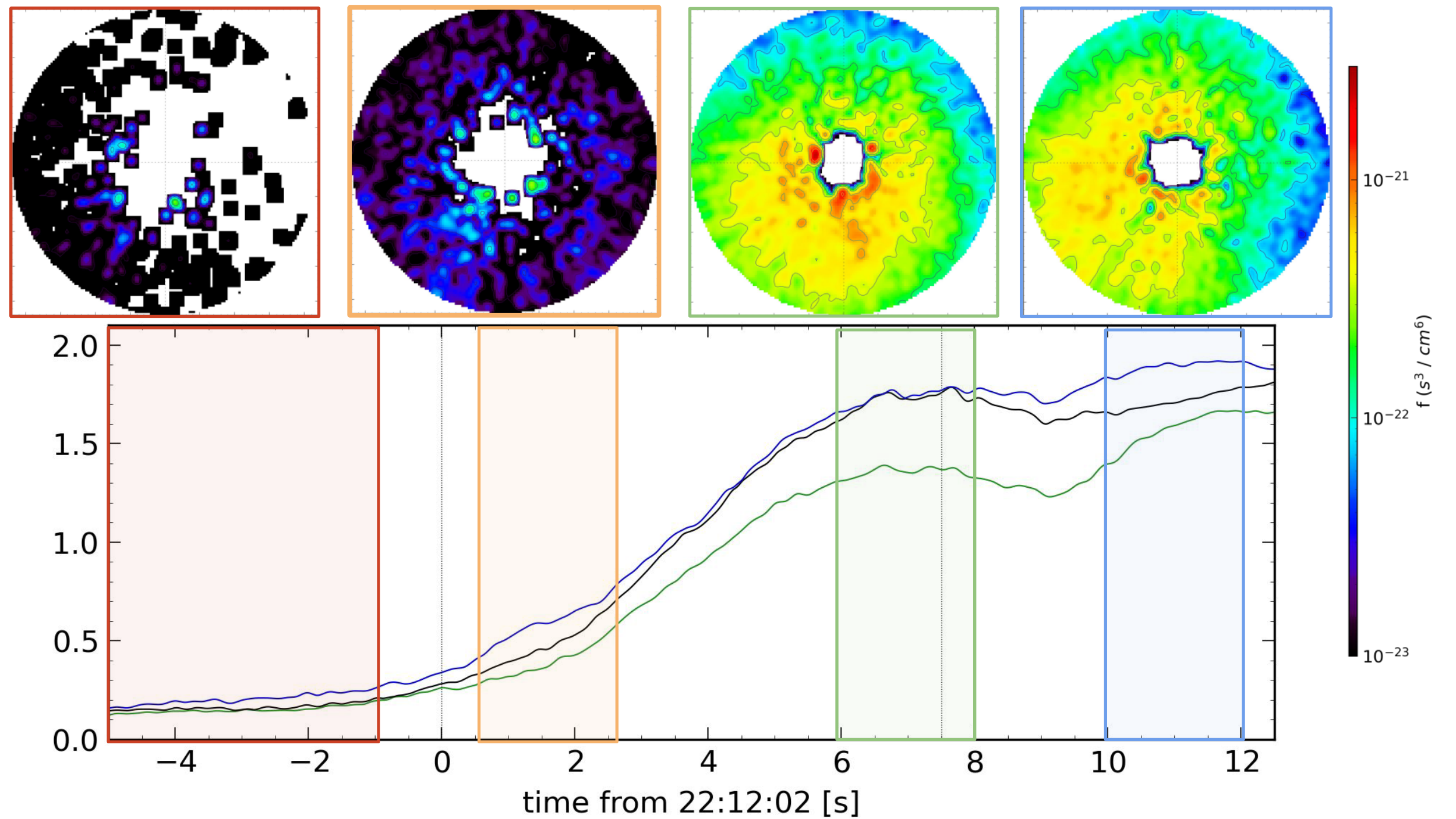}
    \caption{Top: Ions' velocity distribution functions in the tangential plane (the $\mathbf{T}_1$- $\mathbf{T}_2$ plane) averaged in four different time periods.  Velocities axes are between -220 km/s and 220 km/s. Bottom: Eigenvalues of the pressure tensor (same interval as in the red dashed square of Fig. \ref{fig:non-gyrotropy}.c). The four colored boxes are used to distinguish the four time intervals.}
    \label{fig:VDFs}
\end{figure}
Finally, we analyzed the non-gyrotropy with respect to a generic direction, $i.e.$ without imposing that this direction is the magnetic field direction. Specifically, we have looked at a direction, denoted as $\mathbf{g}$, around which the rotated matrix could be rewritten as follows:
\begin{equation}
\begin{pmatrix}
P_2 & 0 & 0\\
0 & P_1 & 0\\
0 & 0 & P_1
\end{pmatrix}
\end{equation}
To achieve this, we employ a minimization algorithm to derive the rotation matrix $\mathbf{M}$ that allows us to put the pressure tensor data under a form as close as possible to this one. Results from this study are shown in Fig. \ref{fig:g_plane} (here shown for MMS2). Panel $a$ displays the variation of $P_1$ and $P_2$ along the crossing, where $P_2$ consistently exceeds $P_1$. In addition, we imposed an upper limit on the temporal variation of the gyrotropic direction $\mathbf{g}$, excluding points with significant temporal variations (indicated by the thin line). Consequently, the remaining points reflect instances where the direction of $ \mathbf{g}$ can be considered as stable and reliable. The vector  $ \mathbf{g} $ itself is represented in panels $b$ and $c$, where it is clear that the direction of gyrotropy is not close to the magnetic field direction: it is close to $\mathbf{n}_{mean}\times\mathbf{B}$, the component along $\mathbf{B}$ being smaller and varying. This result reminds us that at boundaries such as the magnetopause, the strong gradients can break the isotropy as much, and even more here, than the magnetic field, so that gyrotropy can be around another vector than $\mathbf{B}$. A similar remark had already been made in \citet{belmont_kinetic_2012} concerning the modeling of a tangential discontinuity.

\begin{figure}
    \centering
    \includegraphics[width=\columnwidth]{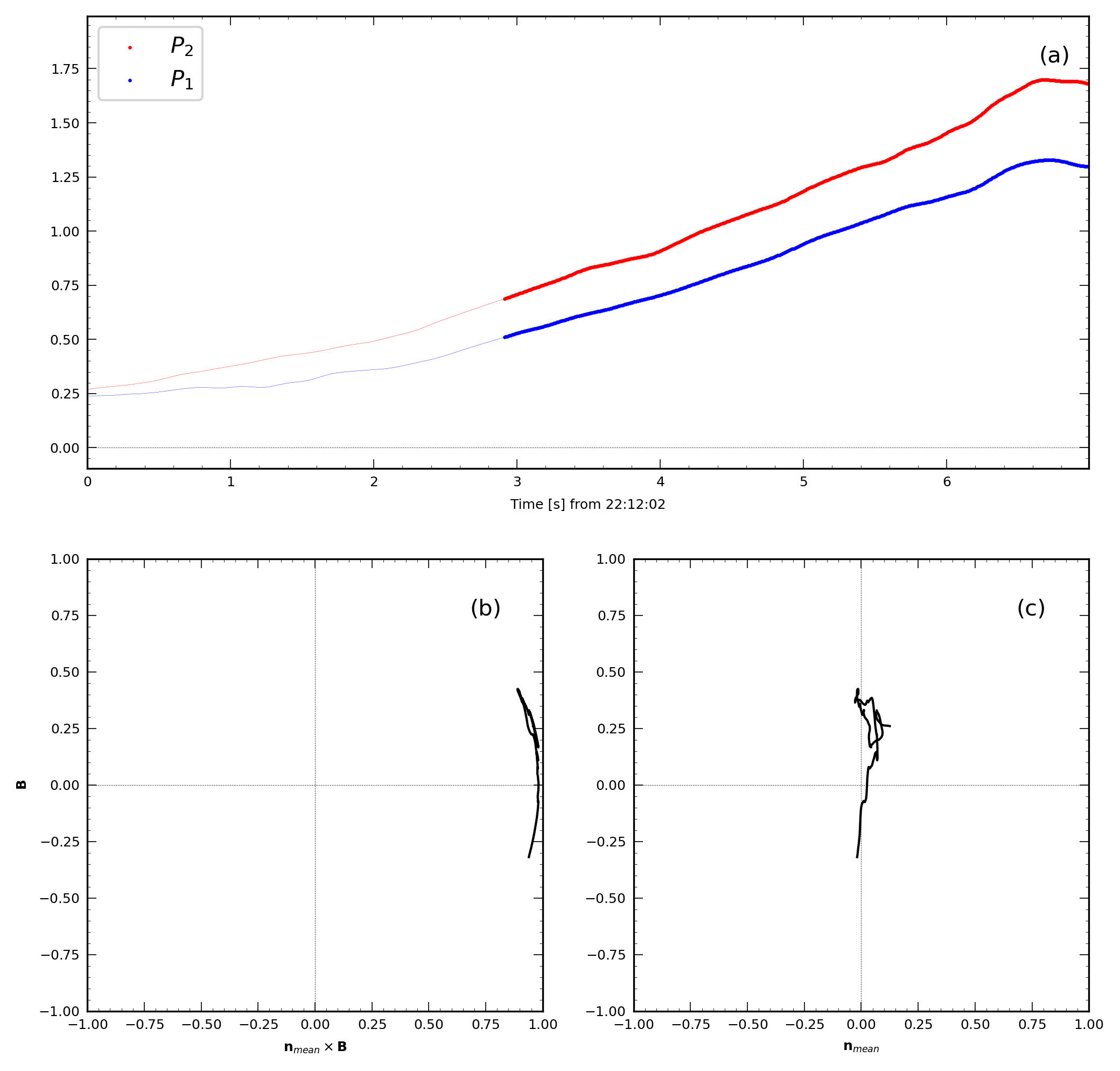}
    \caption{$(a)$: Evolution of parameters $P_1$ and $P_2$. $(b)$ and $(c)$: projections of the gyrotropy direction in two planes. The ordinate is the direction of $\mathbf{B}$, the abscissa is the direction of $\mathbf{n}_{mean}\times\mathbf{B}$ for panel $(b)$ and $\mathbf{n}_{mean}$ for panel $(c)$. }
    \label{fig:g_plane}
\end{figure}

\section{Dataset selection}
In order to expand the results on a statistical basis, we selected a dataset of 146 crossings, chosen from the largest one reported in \citet{nguyen_massive_2022} and \citet{michotte_de_welle_global_2022}. From this database, the following conditions were required in order to carry out an accurate study:
\begin{enumerate}
    \item MMS data are in burst mode.
    \item The crossing duration is between 3 and 15 seconds. Too short crossings do not have a sufficient number of points within the structure (ion measurements are every 0.15 s). Too long crossings may imply non-stationary structures.
    \item Partial crossings are discarded. For that, we impose a density threshold less than 4 cm$^{-3}$ in the magnetosphere and larger than 15 cm$^{-3}$ in the magnetosheath.  
    \item Only cases presenting simultaneous crossing features in particles and magnetic field are considered, in order to compare normals computed at the same time.
\end{enumerate}
In addition to these basic requirements, we also excluded some of the selected crossings for criteria that demand a more detailed analysis of the internal structure of the boundary. First, we excluded two-dimensional features. The quantitative determination of the dimensionality was done with the parameters presented in \citet{Rezeau2017} and the dimensionality index presented in Section \ref{sec:dim}, which are functions of the ratio between the eigenvalues of the gradient matrix. Namely, we considered only crossings with $D_1>0.9$ and $\mathscr{D}_{GF2}>0.8$ , these two parameters being averaged on the crossing interval. These parameters are calculated at each time step but, due to waves and turbulence, attention must be paid that some of these two-dimensional features can be only local and insignificant for the profiles we are looking for. It is the reason why we use only the averaged values. 
\begin{figure}
    \centering
    \includegraphics[width=0.5\columnwidth]{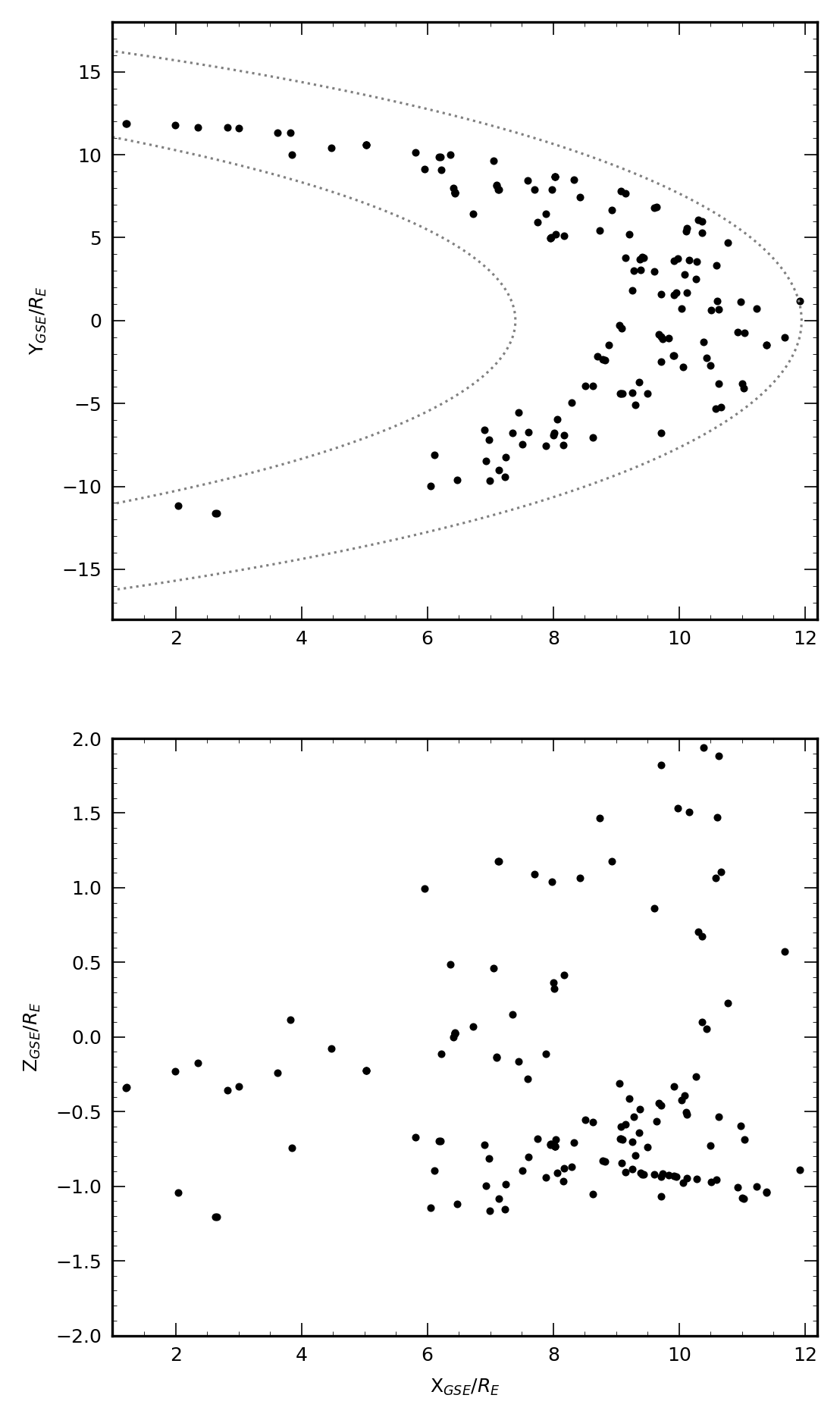}
    \caption{Spatial distribution of the selected database of crossings on the $x,y$ (top) and $x,z$ planes (bottom). The dashed grey lines represent the magnetopause location \citep{shue1997}.}
    \label{fig:positions}
\end{figure}
The 146 selected crossings span from September 2015 to December 2017 (included). We can observe in Figure \ref{fig:positions} that the crossings are evenly distributed in the $x,y$ plane. Regarding the $z$ component, there is a prevalence of cases at negative $z$.

The list of crossings can be found in Supplementary Materials. For each crossing, the classification and the physical quantities relevant for the study (normals, dimensionality index, non-gyrotropy index, and the main characteristic lengths discussed above for the case crossing) are included.

\section{Statistical study of the magnetic hodograms}\label{sec:hodograms}
The previous results about the role of the FLR effects at the magnetopause are now carried out statistically. This study aims to generalize the results obtained from the case crossing studied in the previous section and to estimate the role played by FLRs at the magnetopause.

The database described above has first been used to perform a statistical study on the hodogram shapes, to determine how often linear hodograms are observed in magnetopause crossing. Having an estimation of the percentage of crossings that do not conform to CTD allows us to gauge how frequently the assumptions made by this theory do not accurately represent the magnetopause. For this purpose, we separate the crossings in different classes, this classification being based on CTD distinctions and on the preceding findings:
\begin{enumerate}
    \item \textit{Linear} crossings, $i.e.$ straight lines not passing through the origin as in the above case study.
    \item \textit{Radial} crossings, including all linear crossings whose best fit line passes through the origin (considering uncertainty). These crossings correspond to CTD compressional discontinuities.
    \item \textit{Circular} crossings, when the distance from the origin is constant. These cases correspond to CTD rotational discontinuities.
    \item \textit{Other} crossings, whose features are not included in the previous classes. This class includes crossings with various features, $e.g.$ circular hodograms not centered on the origin, crossing characterised by two different hodograms in two sub-intervals, etc., and  crossings that do not have an obvious distinction between the previous classes, due to noise.
\end{enumerate}

To classify each crossing, we only focus on its central time interval, where the gradients are maximum. By considering larger time intervals, the hodograms' shape becomes more complex because the variations out of this interval are generally unrelated to the main boundary jumps. Selecting only the middle part of the crossing provides simpler and more conformal hodograms. Even if the boundary jumps are not fully completed in this part, this will not prevent comparing the experimental results with CTD predictions since this theory, when valid, is based on conservation laws for any sub-interval of the discontinuity. When this theory fails to reproduce the observed properties, we can interpret those new features as coming from kinetic effects, therefore confirming the limitation of CTD to describe the magnetopause boundary. To that purpose, for each dataset we selected the crossing temporal interval following the algorithm used in \citet{haaland2004, haaland_characteristics_2014} and \citet{paschmann_largescale_2018} to estimate the spatial scale of the magnetopause (intervals are identified as 75$\%$ of the magnetic field $\mathbf{B}_L$ component variation).

The classification performed here differs from previous attempts to classify magnetopause hodograms, as seen in studies such as \citet{sonnerup_magnetopause_1974,berchem_magnetic_1982} and \citet{panov_two_2011}. In these previous works, hodograms were categorized as C-shaped or S-shaped based on their form in the tangential plane. However, unlike those studies, we considered the central part of the crossing, rather than considering the entire temporal interval. Our classification of hodograms involves a two-step process:
\begin{enumerate}
    \item[(1)] Visual Inspection: Initially, all hodograms are visually inspected to identify the cases that are clearly not linear or circular, which are classified separately as 'Others'. Additionally, a preliminary distinction is made between crossings with circular and linear features.
    \item[(2.a)] Analysis of hodograms with possible circular features: For these crossings, we analyze the variation of the modulus of the magnetic field in the plane, allowing for a maximum possible variation of \(20\%\). This accounts for factors such as turbulence and waves propagating alongside the magnetopause. Any crossings exceeding this \(20\%\) threshold are categorized as 'Others.'
    \item[(2.b)] Analysis of hodograms with possible linear or radial features: These crossings undergo an initial assessment to confirm their linearity. This involves examining the width-to-length ratio of the crossing, with any ratio exceeding \(20\%\) classified as 'Other.' Finally, the remaining crossings are classified as either radial or linear based on whether their projection passes through the origin.
\end{enumerate}

From this database, we found the following distribution:
\begin{itemize}
    \item[-] 36.3$\%$ (53/146) of the crossings present linear features.
    \item[-] 2.7$\%$ (4/146) of the crossings present circular features (rotational discontinuity).
    \item[-]15.8$\%$ (23/146) of the crossings present radial features (compressional discontinuity).
    \item[-]45.2$\%$ (66/146) of the crossings could not be interpreted definitely as either of the three before (presenting more than one feature at the same time).
\end{itemize}
It follows that more than a third of the selected crossings show linear features, emphasizing that the fundamental role FLR effects have on magnetopause structure is found in a significant number of crossings.

It could be interesting to compare the above results with the several classifications that were previously published (see \citet{Liu2022} and references therein). These previous classifications were not based on the analysis of the rotational and compressional properties as done here, but on the normal component of the magnetic field and its magnitude (background and variation) ~\citep{smith_identification_1973, burlaga_interplanetary_1977, tsurutani_interplanetary_1979, neugebauer_progress_2010}. For such a comparison, however, one should take care that there are important differences in the definitions: in these previous classifications in particular, any discontinuity is named "tangential", whatever its other properties, as soon as the measured $B_n$ is sufficiently smaller than $B$, the threshold for this ratio being for instance of the order of 0.3 ~\citep{Liu2022,smith_identification_1973, burlaga_interplanetary_1977, tsurutani_interplanetary_1979, neugebauer_progress_2010}. This is of course a very different approach from the one we use here since, even when $B_n$ is small (and even if barely measurable), we consider that different kinds of discontinuities exist, with different properties.

As done for the case study above, it was possible to study on a statistical basis $i)$ the ratio between the width of the magnetopause and the ion Larmor radius and $ii)$ the non-gyrotropy index. For both parameters, the case study appears rather typical. On average, the magnetopause was found to be approximately 6.5 times the ion Larmor radius, only slightly smaller (6.1) for linear hodograms. Similarly, the non-gyrotropy index $D_{ng, \perp}$, has an average value of 0.07, only slightly higher (0.08) for linear hodograms. The $D_{ng, Aunai}$ index has even comparable averages for the four different classes. It therefore seems that, although non gyrotropy has been demonstrated above to play an important role, the non-gyrotropy index alone is not decisive for predicting unequivocally the shape of the hodograms. This question should be the subject of future works.

\section{A comparison between the magnetic and the particles normals}\label{sec:normals}
For each crossing, both the magnetic and the particles normals were computed with the GF2 tool. Thanks to the high resolution of the MMS measurements, we can measure the local fluctuations of the normals inside the magnetopause around their mean values. However, in order to compare the magnetic and ion geometries, a single average normal was used for each case. The mean normal is obtained inside the same time interval as in the previous study. 

To study the differences between the two normals, we compared them via their departure from the Shue model's normal (where the magnetopause is assumed to be a paraboloid, \citep{shue1997}). This normal was obtained using the solar wind and IMF properties from the OMNI data set~\citep{king_solar_2005}. The time delay between the crossing time and the measurement time of the solar wind relevant parameters is estimated by using the propagation method used in \citet{michotte_de_welle_global_2022} (which was adapted from \citet{safrankova_magnetopause_2002}). The procedure for acquiring these parameters is as follows: $i)$ the distance from the bow shock's nose (where OMNI data are defined) to the crossing location, projected along the Earth-Sun axis, is estimated; $ii)$ we estimate the solar wind's propagation time ($t_{est}$) between these two points, assuming an average solar wind velocity of 400 km/s; $iii)$ the solar wind velocity $V_{sw}$ is then determined from the OMNI dataset, averaging over a 2-minute interval centered on the crossing time adjusted by the time delay $t_{est}$; and (IV) ultimately, a final time delay is computed based on $V_{sw}$, which is subsequently utilized to obtain final values of solar wind and IMF parameters. The crossings for which OMNI data computed with this procedure are missing (10 out of 146) were left out of this analysis.
\begin{figure}
    \centering
    \includegraphics[width=0.8\columnwidth]{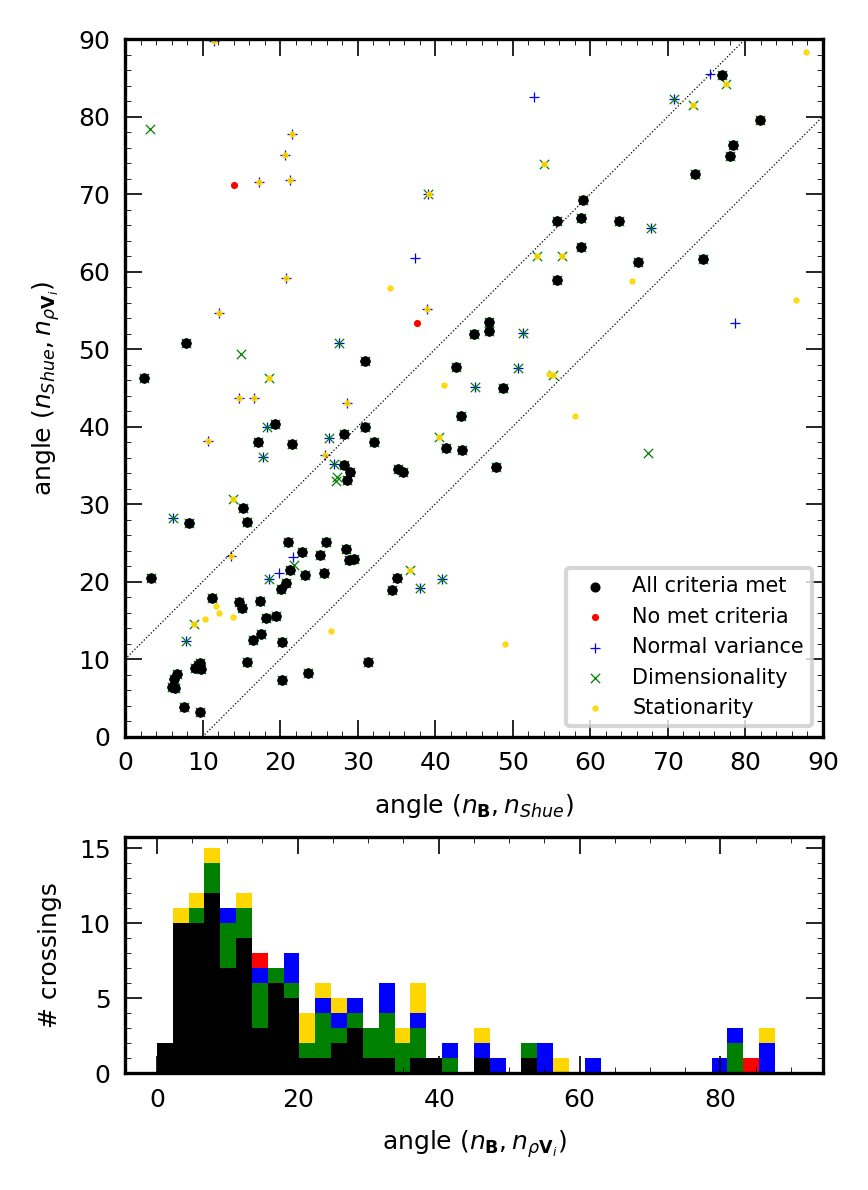}
    \caption{($a$) Comparison between the angle between the theoretical normal \citep{shue1997} and the magnetic and ion normals, ($b$) Distribution of the angle between the magnetic and ion normals. Here the the markers for each point is chosen depending on whether each crossing respects the criteria on dimensionality, stationarity, and normal variance on the ions flux measurements (see Appendix \ref{app:ions} for further details). Colors in the histograms are used accordingly. Blue, green and yellow points indicate the crossing with small variance on the ions normal direction within the crossing, good one-dimensionality and good stationarity. Black points indicate the crossings respecting all criteria, red points not any criteria.}
    \label{fig:statistics}
\end{figure}

In Figure \ref{fig:statistics}$.a$, we plot the angle between the nominal normal and the magnetic and particles normals respectively. In this figure we observe that most of the crossings are along the diagonal, corresponding to cases where the two normals, ionic and magnetic, are similar (82 points out of 146 are between the two thin lines, which indicate differences of $\pm{10^o}$). 

The cases are distributed throughout the plane, with many cases above 40$^o$, although we observe a cluster at lower angles, between zero and 30$^o$. 
The largest angles correspond to a magnetopause very far from the paraboloid shape assumed in Shue's model, which relies on the assumption of a magnetopause at (or near) equilibrium. The departures are likely to be related to surface waves on the boundary itself. 

Finally, the distribution of the angles between the two normals is shown in Fig. \ref{fig:statistics}$.b$. Here we evidence again that most of the cases studied (82 out of 146) are below 20$^o$, with the maximum of the distribution at 10$^o$. However, we also observe again that several cases have much larger angles, up to 90$^o$.
The strongest departures are problematic and deserve further investigation. This appears to be due to the more complex ion structure with respect to the magnetic one. As the criteria used for the dataset selection were built from magnetic data, they are not as relevant when considering ion normals. This is evidenced in Figure \ref{fig:statistics}, where the colors indicate how several ion criteria are satisfied. These criteria concern respectively the dimensionality, the stationarity, and the variance of the normal direction. All details are given in Appendix \ref{app:ions}. Focusing on points respecting all the criteria for the ions flux (black markers and hodogram), we observe that only a few crossings are outside the diagonal. Only two of these crossings have angles above 40$^o$.

\section{Conclusions}
The study of the properties of the magnetopause is a very important issue for understanding the penetration of the solar wind plasma into the magnetosphere. In the theoretical part, we show that the notion of "quasi-tangential" discontinuity has to be introduced to complete the theory of discontinuities and understand the limit when the crossing fluxes tend to zero as in the magnetopause case. We emphasize that, in presence of anisotropy, the physical processes occurring inside the layer play a fundamental role  because they are responsible for the conditions linking the downstream and upstream quantities. In particular, for thin current layers, the FLR corrections corresponding to the non-gyrotropic pressure tensor components must be taken into account.

The tool GF2 presented in the paper and used for determining the normal direction to the boundary derives from the MDD method. It includes in addition a fitting procedure, which allows introducing a part of the temporal information via a 4-point filtering of the data and adding constraints such as $\mathbf{\nabla} \cdot \mathbf{B}=0$. It is shown here to provide results quite compatible with the original method (when used with smoothed data), which is enough for drawing reliable physical conclusions on the magnetopause equilibrium. We expect that this approach could bring more precise information concerning the magnetopause gradients. Unfortunately, investigating this point in more detail cannot be done using MMS data but requires testing the tool in fully 3D kinetic simulations with realistic turbulence. This point is the subject of future work. Here, we have applied this tool on a particular crossing case and compared with other state-of-the-art normals. We have shown that the local normal (at each time step during the crossing) differs by less than ten degrees from the one calculated by all the other models. When averaging over the whole crossing, the normal obtained with the GF2 is even less than one degree apart from the normals from \citet{shi_dimensional_2005, Denton2018}.

Although we cannot claim to have achieved the ideal accuracy of about one degree, the reached accuracy is sufficient to evidence the correct physics at play, resumed as FLR effects. We have presented the results for a crossing observed by the four MMS spacecraft. For this crossing, the "linear" hodogram in the tangential plane shows that the boundary properties differ from those predicted by CTD. This discrepancy is explained by looking at the tangential components of the momentum equation, which highlights the role of the pressure tensor symmetries in the magnetopause equilibrium. This result agrees with the theoretical results of the first part and it is likely to hold more generally for all quasi-tangential discontinuities. The ion pressure tensor has been analyzed for this purpose. We have used two indices of non-gyrotropy, which both confirm the presence of a significant, even if small, non-gyrotropic part in this tensor. Furthermore, we have shown that the non-gyrotropy direction differs from the magnetic field one, aligning approximately with the $\mathbf{n}_{mean}\times\mathbf{B}$ direction. Finally, the analysis of the VDFs directly confirms the presence of non-gyrotropic distributions.

To show that our methodology applies to cases that CTD cannot handle, we have selected a substantial number of magnetopause crossings with one-dimensional characteristics to have a proper statistical basis for our findings. For all these crossings, we have plotted the hodogram of the magnetic field in the tangential plane and classified them depending on their geometry. Our results show that 36.3$\%$ of the crossings evidence clear linear features, incompatible with the CTD description, while only 18.5$\%$ of the crossings show either circular or radial hodograms as predicted by CTD. In other words, a significant number of cases escapes the classic theory, proving that the relevance, even if not a predominance, of FLR effects at the magnetopause can be generalized and that the case crossing presented in the first section is rather typical. It is well-known that the linear version of the rotational discontinuity is the MHD shear Alfv\'en wave. Here it appears that the magnetopause-like "quasi-tangential" discontinuities correspond in the same way to the quasi-perpendicular "Kinetic Alfv\'en Waves"~\citep{Hasegawa,belmont_rezeau1987,cramer}.

Several papers have investigated the changes in rotational discontinuities when various non-ideal effects are introduced. These theoretical papers have addressed the problem as a Riemann problem using the methodology of a "piston" to study the formation of different discontinuities. Some introduced FLRs and gyroviscosity in the layer while assuming isotropy on both sides \citep{lyu_1989,hau_1991}, and others introduced anisotropy everywhere while assuming gyrotropy in the layer \citep{hau_2016}. These different papers lead to different conclusions; in particular concerning the role of electron inertia in the layer equilibrium.\\
It is worth noticing that the hodograms of \textbf{B} obtained with these theoretical studies were never far from circular ones, contrary to the almost linear shapes shown in the present paper. Our methodology has been different here: without assuming pre-defined forms for the non-ideal terms, we look experimentally to the hodograms and the form of the \textbf{P} tensor and explain theoretically how the second can explain the first ones. 

Finally, we have used the same database of crossings to compare the geometric properties of the magnetic and ion structures. We have compared the normal obtained from the magnetic field and the ion flux measurements to the one expected from \citet{shue1997} model. Many crossings differ by more than 40 degrees from the nominal equilibrium condition, underlining a very dynamical environment, but it is worth noticing that the two kinds of determination are most often in agreement with each other, and therefore confirm the result. Furthermore, an accurate study of the ion flux measurements have shown that crossings showing bigger discrepancies between the magnetic field and ion flux normals are generally due to non-stationarities, non-one-dimensionality, or variations in the ion flux normals. When excluding these cases from the study, the ion and magnetic flux normals are compatible with only two crossings (over 77) showing angles larger than forty degrees. 

\section*{Data availability}
Magnetospheric  Multiscale  satellite  data  were  accessed  through  the  MMS  Science  Data  Center, \url{https://lasp.colorado.edu/mms/sdc/public/}. Furthermore, all the softwares employed, from interpolation of the data to the analysis itself,  can be found at \url{https://github.com/GiulioBallerini/Notebooks_FLR.git}.

\begin{acknowledgements}
\noindent \textbf{Acknowledgements:} The authors strongly thank Nicolas Aunai and Bayane Michotte de Welle for useful discussions. The French involvement on MMS is supported by CNES and CNRS.\\
\textbf{Declaration of interests:} The authors report no conflict of interest.\\
\textbf{Funding:} The first author, Giulio Ballerini, acknowledges the support of "Ecole Franco-Italienne' (grant number C2-222)
\end{acknowledgements}
\appendix
\section{Influence of averaging the $\mathbf{t}_1$ direction in the momentum equation balance}
\begin{figure}
\centering
\includegraphics[width=0.5\textwidth]{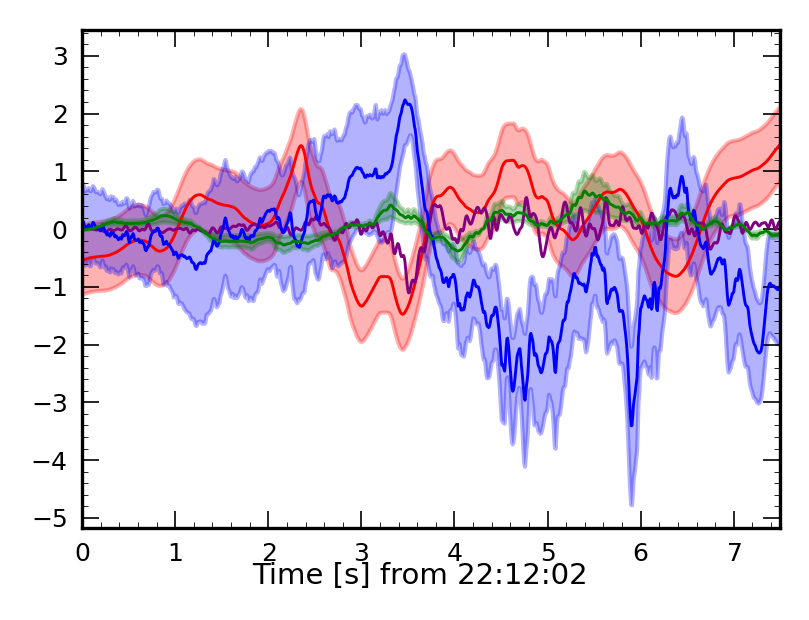}
\caption{\label{fig:appendixA} Terms of the momentum equation (units of $10^{-15}kg\,m/s^2$), projected on the local tangential direction ($\textbf t_1$).  Shaded regions are estimated uncertainties of the divergence of the pressure (red), the $\mathbf{J}\times\mathbf{B}$ (blue) and the classic inertial term (green).}
\end{figure}
\label{appendix}
In this section, we investigate the impact of using an averaged tangential direction along the crossing on the outcomes concerning the role of the pressure tensor in the momentum equation. In Fig. \ref{fig:appendixA} we show the projection of the terms of the momentum equation along the local $\mathbf{t}_1$ direction ($i.e.$ without averaging). We observe here some reversals of the sign of the dominant terms, that were not observed in the averaged case. Nonetheless, it is still evident that the pressure tensor counterbalances the $\mathbf{J}\times\mathbf{B}$ term, hereby confirming our earlier findings.

\section{Analysis of the gyroviscous effects}\label{app:gyroviscous}
In this section, we use the magnetopause crossing analysed in detail above to study the validity of the gyroviscous interpretation. In particular, we employ the Braginskii gyroviscosity term \citep{braginskii_transport_1965} as applied by \citet{Stasiewicz1989} to the magnetopause, to analyze the pressure tensor. In this case, the pressure tensor is considered as the sum of an isotropic component, $\mathbf{P}_{iso}$ and a viscosity term, $\mathbf{\sigma}$:
\begin{equation}
    \mathbf{P}_{i}=\mathbf{P}_{iso}-\mathbf{\sigma}
\end{equation}
To investigate the viscosity term, we use the reference system where the normal direction is aligned with the z-axis (the x and y directions are chosen accordingly to form an orthogonal triad). By exploiting the definition of $\mathbf{\sigma}$, we focus here on its projection along the normal yielding the following relation:
\begin{equation}
-\mathbf{\sigma}.\mathbf{n}=
\begin{pmatrix}
P_{nx}\\
P_{ny}\\
P_{nn}
\end{pmatrix}
=\rho\nu
\begin{pmatrix}
0 & b_n & b_y\\
-b_n & 0 & -b_x\\
b_y & -b_x & 0
\end{pmatrix}
.\begin{pmatrix}
u_{x}'\\
u_{y}'\\
u_{n}'
\end{pmatrix}
\end{equation}
Here $\nu$ is the gyroviscosity coefficient, $\mathbf{\hat{b}}=(b_x, b_y, b_z)$ the normalized magnetic field, and $\mathbf{u}'=(u_x', u_y', u_z')$ is the vector of the spatial derivatives of the velocity components along the normal. We now consider the first two components of this equation, yielding the following expressions that allow us to compare the non-diagonal terms with the velocity changes:
\begin{equation}
    \frac{P_{nx}}{\rho}=\nu(b_n u_y'+b_yu_n')   
    \label{eq:gyro1}
\end{equation}
\begin{equation}
    \frac{P_{ny}}{\rho}=-\nu(b_n u_x'+b_xu_n')
    \label{eq:gyro2}
\end{equation}
The terms of these equations are shown (normalized) in Figure \ref{fig:gyroviscous}. Here we observe a fairly good correlation between the non-diagonal terms of the pressure tensor and the spatial derivatives of the flow velocity.

\begin{figure}
    \centering
    \includegraphics[width=0.85\columnwidth]{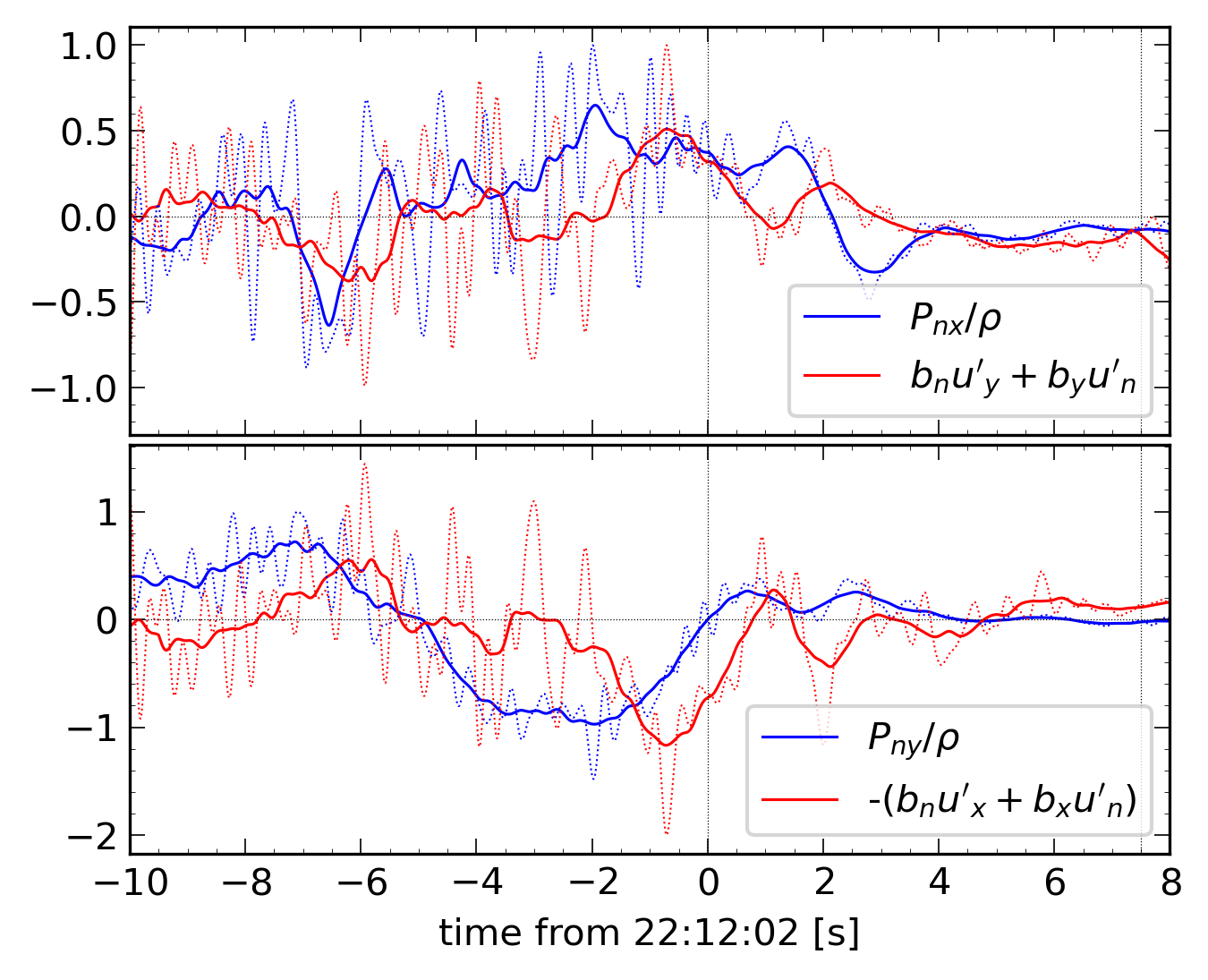}
    \caption{Left (blue) and right (red) hand sides for Equations \ref{eq:gyro1} (top) and \ref{eq:gyro2} (bottom). Thin-dotted lines correspond to the real-time values while thick lines to an averaged window of 1 $s$. All terms are normalized.}
    \label{fig:gyroviscous}
\end{figure}

\section{Quality indices for the ion normals} \label{app:ions}
In the absence of additional caution, Figure \ref{fig:statistics} shows that the angle between the normal obtained with the magnetic field and the one with the ion flux reaches very high values, up to 90 degrees. This result requires a more accurate study, as the criteria used for the dataset selection are based on the magnetic field (except for the threshold imposed on the density values).\\
To interpret the results accurately, the following parameters were considered:
\begin{enumerate}
	\item  Dimensionality of ion flux. For this purpose, we exploit the dimensionality index defined in Equation \ref{eq:dim}, computed from the ion flux measurements.
	\item Stationarity of the ion flux measurements. To evaluate stationarity, we exploit the GF2 tool. Specifically, we consider the quality of the fit of the gradient matrix as an index of stationarity. By defining $\mathbf{D} = \mathbf{G}_{\text{fit}} - \mathbf{G} $ we can introduce the stationarity index:
	\begin{equation}
	    S=\frac{\Tr (\mathbf{D}. \mathbf{D}^T)}{\Tr (\mathbf{G}. \mathbf{G}^T)}
	\end{equation}
	Since for a truly stationary magnetopause, $S$ should be equal to zero, deviations from zero suggest potential non-stationarity.
	\item Variance of the normal. In some crossings of the database, the normal associated with ion flux exhibits local differences with respect to the mean value, such as fluctuations or rotations within a plane, with one component varying within the crossing. In these cases, the ion flux is therefore characterized by more complex structures and the mean normal is not meaningful. To exclude such cases, we examined the variation of the normal around the mean value, defined as follows:
	\begin{equation}
	    \delta_{norm}=< |\mathbf{n}_i-\mathbf{n}_{mean,i}|^2 >
	\end{equation}
    Small values of $\delta_{norm}$ indicate almost constant normals.
\end{enumerate}
The average values of these three parameters for each crossing are shown in Figure \ref{fig:ions_study} as a function of the angles between the normal of the magnetic field and the ion flux. We observe here that crossings showing the largest angles occur when at least one of these conditions fails. To select the cases for which the ions are characterized by a stationary and one-dimensional structure, for which the normal has no variations around the mean value, we applied the following thresholds: $\mathscr{D}_{GF2, ions}>0.6$, $\delta_{norm}>0.07$, $S>0.22$. Specifically, crossings individually meeting one of these criteria are shown in green, blue, and yellow, respectively. When all criteria are met, crossings are indicated by black dots.  This Figure underlines a correlation between the difference between the two normals and the values of these three parameters, showing how cases with higher $\mathscr{D}_{GF2, ions}$ and smaller $\delta_{norm}$ and $S$ are the ones with smaller differences between the two normals.

\begin{figure}
    \centering
    \includegraphics[width=\columnwidth]{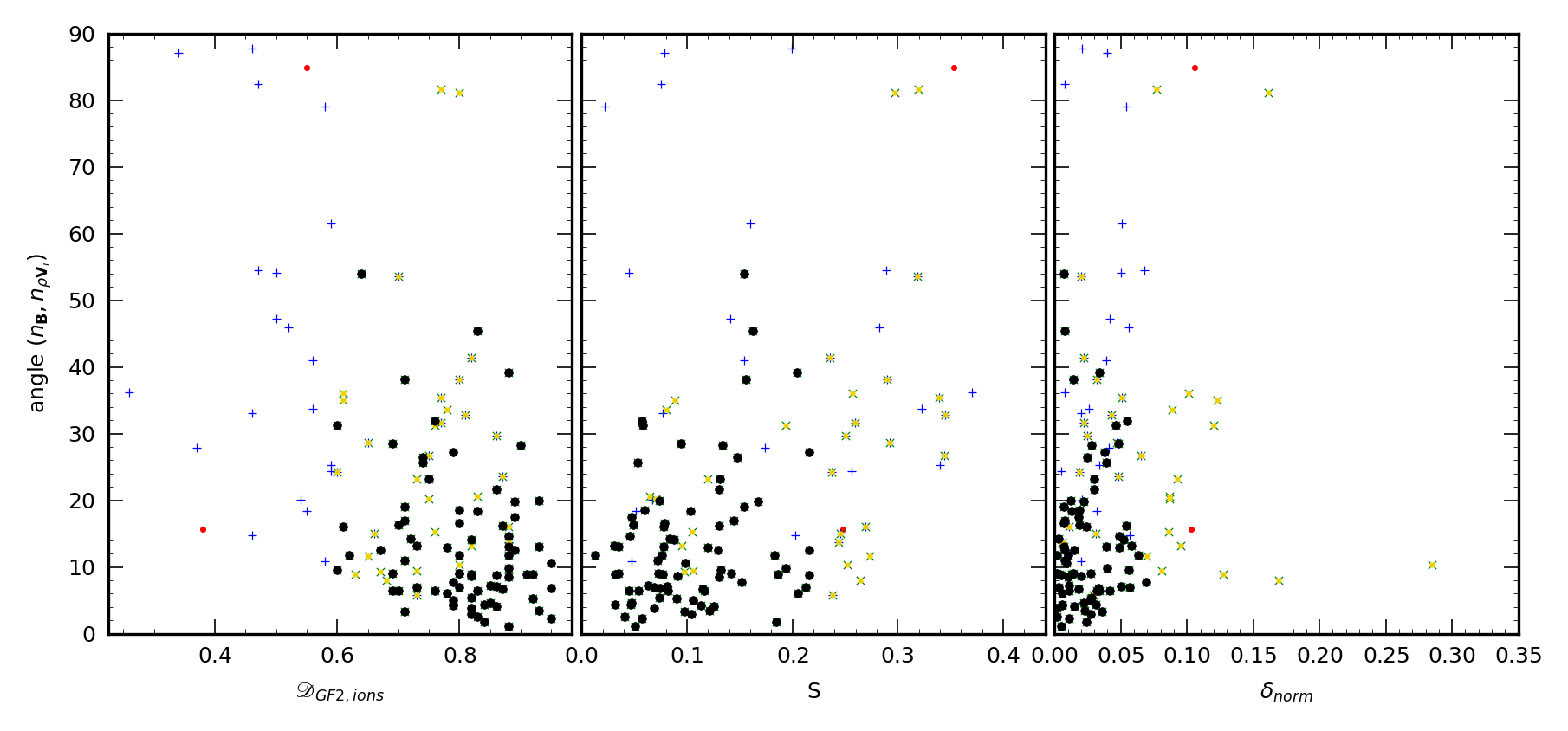}
    \caption{Dimensionality (left), stationarity (center), and normal variance (right) averaged for each crossing as a function of the angle between the magnetic field normal and the ion flux one. Green, blue, and yellow indicate crossings respecting the $\mathscr{D}_{GF2, ions}>0.6$, $\delta_{norm}>0.07$, $S>0.22$ criteria individually. Black dots indicate the crossings for which all the criteria are met, and red dots (two cases) when no condition is met.}
    \label{fig:ions_study}
\end{figure}}
\bibliographystyle{jpp} 
\bibliography{biblio}

\end{document}